\documentclass[aps,prl,reprint,superscriptaddress]{revtex4-2}

\pdfoutput=1
\usepackage{amsmath,amssymb,bm}
\usepackage{color,xcolor}
\usepackage{graphicx,epsfig}

\usepackage{microtype}
\usepackage{hyperref}
\usepackage[capitalize]{cleveref}
\hypersetup{
  colorlinks=true,
  citecolor=teal,
  linkcolor=teal,
  urlcolor=cyan,
  filecolor=blue,
  breaklinks=true,
}
\begin{document}

\title{Anomalous grain dynamics and grain locomotion of odd crystals}

\author{Zhi-Feng Huang}
\affiliation{Department of Physics and Astronomy, Wayne State University, Detroit, Michigan 48201, USA}
\author{Michael te Vrugt}
\affiliation{Institut f\"{u}r Physik, Johannes Gutenberg-Universit\"{a}t Mainz, 55128 Mainz, Germany}
\author{Raphael Wittkowski}
\affiliation{Department of Physics, RWTH Aachen University, 52074 Aachen, Germany}
\affiliation{DWI -- Leibniz Institute for Interactive Materials, 52074 Aachen, Germany}
\author{Hartmut L\"{o}wen}
\affiliation{Institut f\"{u}r Theoretische Physik II: Weiche Materie, Heinrich-Heine-Universit\"{a}t D\"{u}sseldorf, 40225 D\"{u}sseldorf, Germany}

\begin{abstract}
Crystalline or polycrystalline systems governed by odd elastic responses are known to exhibit complex dynamical behaviors involving self-propelled dynamics of topological defects with spontaneous self-rotation of chiral crystallites. Unveiling and controlling the underlying mechanisms require studies across multiple scales. We develop such a type of approach that bridges between microscopic and mesoscopic scales, in the form of a phase field crystal model incorporating transverse interactions. This continuum density field theory features two-dimensional parity symmetry breaking and odd elasticity, and generates a variety of interesting phenomena that agree well with recent experiments and particle-based simulations of active and living chiral crystals, including self-rotating crystallites, dislocation self-propulsion and proliferation, and fragmentation in polycrystals. We identify a distinct type of surface cusp instability induced by self-generated surface odd stress that results in self-fission of single-crystalline grains. This mechanism is pivotal for the occurrence of various anomalous grain dynamics for odd crystals, particularly the predictions of a transition from normal to reverse Ostwald ripening for self-rotating odd grains, and a transition from grain coarsening to grain self-fragmentation in the dynamical polycrystalline state with an increase of transverse interaction strength. We also demonstrate that the single-grain dynamics can be maneuvered through the variation of interparticle transverse interactions. This allows to steer the desired pathway of grain locomotion and to control the transition between grain self-rotation, self-rolling, and self-translation. Our results provide insights for the design and control of structural and dynamical properties of active odd elastic materials.
\end{abstract}

\maketitle

The understanding and control of grain growth dynamics are among the key subjects in the study of crystalline and polycrystalline materials. Typical phenomena include the classical process of Ostwald ripening \cite{Ratke02}, which corresponds to the growth of larger grains at the expense of vanishing smaller ones via intergrain particle diffusion through the liquid medium, and the dynamics of grain coarsening in polycrystals, for which the evolution of topological defects, particularly grain boundaries and dislocations, plays a crucial role \cite{CahnActaMater04,QiuScience24}. Most of the conventional passive materials are potential systems, where microscopically the interparticle interactions are governed by center-center longitudinal forces and the corresponding interaction potentials, while mesoscopically and macroscopically the system evolution and grain dynamics are driven by the pathways towards the equilibrium state characterized by minimum free energy of the system. However, this principle of thermodynamic-potential minimization can no longer be applied for active systems, particularly when the microscopic interactions involve nonconservative transverse forces oriented perpendicularly to the center-center direction of the two interacting active bodies (as originating from, e.g., hydrodynamic near-field interactions between spinning particles \cite{LeoniEPL10,FilySoftMatter12,PetroffPRL15,TanNature22,BililignNatPhys22}). This can lead to the existence of nonreciprocal torques between two active particles; the resulting many-body systems are nonpotential and intrinsically nonequilibrium.

Although in the bulk state of a perfect crystalline lattice, all the interparticle transverse forces are balanced and do not affect the crystal stability, small deformations will lead to odd elastic responses, as identified in continuum odd elasticity theory \cite{ScheibnerNatPhys20}, with antisymmetric or odd elastic constant tensor $C_{ijkl}^{\rm (o)} = -C_{klij}^{\rm (o)}$ and asymmetric stress $\sigma_{ij} \neq \sigma_{ji}$ showing nonzero internal torque caused by transverse forces and the lack of angular momentum conservation. This results in self-generated dynamics of topological defects inside crystalline grains and the self-rotation of odd crystallites induced by nonzero net odd stress and torque on surface, as observed in both experiments \cite{PetroffPRL15,TanNature22,BililignNatPhys22} and simulations \cite{PetroffPRL15,TanNature22,BililignNatPhys22,BravermanPRL21,PoncetPRL22,Choi24} of various active and living chiral crystals formed by, e.g., swimming bacteria \cite{PetroffPRL15}, starfish embryos \cite{TanNature22}, and magnetic colloids \cite{BililignNatPhys22}. More complex behaviors occur for large-scale odd polygrain systems activated by persistent dynamics of motile defects, such as the self-kneading polycrystal whorl state found in recent magnetic colloidal experiments \cite{BililignNatPhys22}.

Such chiral odd crystalline systems naturally involve multiple spatial and temporal scales that mutually couple, in particular microscopic scales at the individual particle or discrete-lattice level and mesoscopic scales of structural patterns or interfacial profiles across, e.g., solid-liquid or grain boundaries. However, current theoretical and computational studies of odd elastic systems mostly focus on two ends of the scaling spectrum, i.e., discrete particle-based simulations \cite{PetroffPRL15,TanNature22,BililignNatPhys22,BravermanPRL21,PoncetPRL22,Choi24,CaporussoPRL24,CapriniM2025}, by which only limited length and time scales can be accessed, and continuum elasticity theory \cite{ScheibnerNatPhys20,BravermanPRL21,ShankarNatPhys24} for the description of long-wavelength behaviors of the system but with the absence of microscopic crystalline symmetry and structures. Although continuum field theories have been developed for the study of nonreciprocal active matter, including nonreciprocal coupling between different species \cite{BowickPRX22,YouPNAS20,BraunsPRX24,SahaAG2020,PisegnaPNAS24,GrevePRL25} and nonreciprocal interactions within a single species \cite{Huang24,KoleCMHRL2025}, what has been still lacking is a truly multiscale approach which incorporates the coupling and bridging between microscopic/discrete and mesoscopic/continuum scales for the modeling of odd crystalline or polycrystalline systems. This hinders our efforts on further examining and predicting the complex dynamical behavior of odd grains with elastic or plastic deformations.

Here we develop such a micro- and meso-scale approach based on the phase field crystal (PFC) method, which incorporates discrete crystalline structure into continuum density field theory and well captures the system elasticity, plasticity, and multigrain dynamics \cite{ElderPRL02,ElderPRB07,Provatas10,ChanPRE09,MkhontaPRL13,WangPRB18,SalvalaglioMSMSE2022}, as has been demonstrated through its wide range of applications in solid, soft, and active systems \cite{EmmerichAdvPhys12,AdlandPRL13,BackofenActaMater14,MoatsPRE19,Salvalaglio20,LiuPRM22,BurnsActaMater24,MenzelPRL13,HuangPRL20,HuangCP22,Holl24}. Effects of transverse interactions are introduced into this continuum field model as nonpotential contributions through derivation from microscopic particle dynamics, leading to a PFC model incorporating transverse interactions (T-PFC model) and the corresponding mesoscale amplitude equation formulation. The modeling reveals a number of intriguing properties of two-dimensional (2D) odd elastic crystals and grains involving complex dynamics of topological defects, with the underlying mechanisms elucidated via the development of a distinct type of surface cusp instability that leads to grain self-fission through dislocation proliferation. The model predictions also include three types of transitions for odd grain dynamics, namely a transition from normal to reverse Ostwald ripening, a transition between the traditional behavior of grain coarsening and the persistent grain self-fragmentation in the dynamical polycrystalline state, and in terms of grain locomotion, the controlled transition from self-rotation to self-rolling (a combination of self-rotation and translation) and to self-translation of grains. Our results are not only consistent with existing experiments and particle-based or molecular dynamics (MD) simulations of various active and living chiral crystalline systems, but also predict mechanisms and conditions that are applicable and extendable to the further study of chiral material systems with odd elastic response.

\section*{Model}
\label{sec:model}

Based on the PFC approach, we represent the spatiotemporal distribution of particles by a local particle density variation field $\psi(\mathbf{r},t)$. Contributions from longitudinal interactions are described by the PFC free energy functional $F_{\rm PFC} = \int d\mathbf{r}\, \{ \frac{1}{2} \psi [-\epsilon + (\nabla^2 + q_0^2)^2 ] \psi - \frac{1}{3} g \psi^3 + \frac{1}{4} \psi^4 \}$. This represents a minimal model for crystal formation, which depending on its parameters produces either a homogeneous (liquid) or a spatially ordered (crystalline) phase \cite{Provatas10,EmmerichAdvPhys12}. The parameter $\epsilon$ is a temperature-like parameter that determines whether the system is above or below the freezing transition, and $q_0$ determines the periodicity of the crystalline phase. We here set $q_0=1$ after rescaling by choosing the lattice spacing of the crystalline phase as a length scale. From a statistical mechanics point of view, the rescaled model parameters $\epsilon$ and $g$ can be expressed via the Fourier components of interparticle direct correlation functions \cite{ElderPRB07,HuangPRE10}. Specifically, the first two quadratic terms in $F_{\rm PFC}$ are related to the two-point direct correlation function, with $\epsilon$ measuring the distance from the crystal melting point and the second gradient term giving the spatially periodic structure of characteristic lattice wave number $q_0$. The cubic and quartic terms can be viewed as those from a Landau expansion which, as combined with the quadratic terms, give rise to the first-order liquid-solid phase transition. They have also been shown to connect to the three-point direct correlation function in classical dynamical density functional theory \cite{HuangPRE10}.
The corresponding flux is given by $\mathbf{J}_{\rm PFC} = -\bm{\nabla} \mu_{\rm PFC} = -\bm{\nabla} \delta F_{\rm PFC}/\delta\psi$, where $\mu_{\rm PFC}$ is the chemical potential. The transition and coexistence between solid and liquid (homogeneous) states are determined by $\epsilon$ and the average density $\bar{\psi}_0$. 

For the odd solid system studied here, we incorporate the nonpotential contribution by coarse-graining the microscopic equations of motion of particles interacting via a transverse force $\mathbf{F}_{ab}^\perp = f^\perp(\mathbf{r}_a,\mathbf{r}_b) \hat{\mathbf r}_{ab}^\perp$ between any two particles $a$ and $b$, with $\hat{\mathbf r}_{ab}^\perp= \hat{\mathbf z} \times \hat{\mathbf r}_{ab}$, $\hat{\mathbf z}$ the out-of-plane unit vector, and $\hat{\mathbf r}_{ab} = (\mathbf{r}_a - \mathbf{r}_b)/|\mathbf{r}_a - \mathbf{r}_b|$. Details of the derivation are given in SI Appendix. The nonpotential flux is found to be 
\begin{equation}
    J_{\mathrm{T},i} = -\psi \left ( \alpha_0 + \alpha_1 \nabla^2 + \alpha_2 \nabla^4 \right ) \epsilon_{ij} \partial_j \psi,
\label{eq:transversecurrent}
\end{equation}
where $i,j=x,y$, $\epsilon_{ij}$ is the 2D Levi-Civita symbol, and $\alpha_k$ ($k=0,1,2$) are proportional to the strength of transverse interaction (with expressions determined by the specific form of the transverse force $f^\perp$). Here the coefficients $\alpha_k$ are allowed to vary with space and/or time. The term proportional to $\alpha_1$ is consistent with the one found in the chiral current of a different type of chiral active matter, namely circle swimmers \cite{BickmannBJW2020}.

The dynamics of the density field is then governed by $\partial \psi / \partial t = -\bm{\nabla} \cdot \mathbf{J} = -\bm{\nabla} \cdot ( \mathbf{J}_{\rm T} + \mathbf{J}_{\rm PFC} )$ with a diffusive timescale, giving the T-PFC model equation
\begin{align}
    \frac{\partial \psi}{\partial t} =&~ \left [ \bm{\nabla} \left (\alpha_0 \psi \right ) \times \bm{\nabla} \psi + \bm{\nabla} \left (\alpha_1 \psi \right ) \times \bm{\nabla} \nabla^2 \psi \right. \nonumber\\
    & \left. ~+ \bm{\nabla} \left (\alpha_2 \psi \right ) \times \bm{\nabla} \nabla^4 \psi \right ]_z + \nabla^2 \frac{\delta F_{\rm PFC}}{\delta \psi}.
    \label{eq:T-PFC}
\end{align}
For spatially and temporally constant coefficients $\alpha_k$,
the T-PFC equation \ref{eq:T-PFC} reduces to
\begin{align}
    \frac{\partial \psi}{\partial t} =&~ \left [ \left ( \bm{\nabla} \psi \right ) \times \bm{\nabla} \left ( \alpha_1 \nabla^2 \psi + \alpha_2 \nabla^4 \psi \right ) \right ]_z \nonumber\\
    & + \nabla^2  \big[ -\epsilon \psi + \left ( \nabla^2 + q_0^2 \right )^2 \psi - g\psi^2 + \psi^3  \big].
    \label{eq:TPFC}
\end{align}
The first term of Eq.~\ref{eq:TPFC}, originating from transverse interactions, breaks the 2D parity symmetry, which can then lead to 2D chirality. It is also antisymmetric with respect to the exchange of $x$ and $y$, resulting in a non-Hermitian dynamical matrix for the elastodynamics of the displacement field, a characteristic of nonreciprocity (SI Appendix). In addition, it is straightforward to prove that this 2D T-PFC equation is rotationally invariant with respect to a rotation about the $z$ axis, i.e., maintaining global rotational invariance in 2D, in addition to translational invariance.

\subsection*{Amplitude Expansion}

The mesoscale description of the model is obtained from the corresponding amplitude expansion, i.e.,
\begin{equation}
    \psi = \psi_0 + \sum_{j=1}^3 A_j e^{i\mathbf{q}_j^0 \cdot \mathbf{r}} + \textrm{c.c.}, \label{eq:Aj_expan}
\end{equation}
where $\mathbf{q}_1^0 = \tilde{q}_0 ( -\frac{\sqrt{3}}{2}\hat{\mathbf x} - \frac{1}{2}\hat{\mathbf y} )$, $\mathbf{q}_2^0 = \tilde{q}_0 \hat{\mathbf y}$, and $\mathbf{q}_3^0 = \tilde{q}_0 ( \frac{\sqrt{3}}{2}\hat{\mathbf x} - \frac{1}{2}\hat{\mathbf y} )$ are the basic wave vectors of a 2D hexagonal lattice, with $\hat{\mathbf x}$ and $\hat{\mathbf y}$ the unit vectors along the $x$ and $y$ directions respectively, and $\tilde{q}_0$ ($\sim q_0$) is the steady-state selected wave number. The 0th-mode average density field $\psi_0$ and the complex amplitude $A_j$ ($j=1,2,3$) vary at the mesoscopic ``slow'' scales. Substituting Eq.~\ref{eq:Aj_expan} into Eq.~\ref{eq:TPFC} and following the procedure of amplitude equation formulation \cite{GoldenfeldPRE05,HuangPRE10} by separating microscopic and mesoscopic scales, we get
\begin{align}
    \frac{\partial \psi_0}{\partial t} =&~ \left [ \left ( \bm{\nabla} \psi_0 \right ) \times \bm{\nabla} \left ( \alpha_1 \nabla^2 \psi_0 + \alpha_2 \nabla^4 \psi_0 \right ) \right ]_z \nonumber\\
    &~ + \sum_j \left \{ \left [ \left ( \bm{\nabla} + i\mathbf{q}_j^0 \right ) A_j \right ] \times \left ( \bm{\nabla} - i\mathbf{q}_j^0 \right ) \right \}_z \nonumber\\
    &\qquad \mathcal{L}_{j}^* \left ( \alpha_1 + \alpha_2 \mathcal{L}_{j}^* \right ) A_j^* + \textrm{c.c.} + \nabla^2 \frac{\delta\mathcal{F}_{\rm PFC}}{\delta\psi_0}, \label{eq:psi0}\\
    \frac{\partial A_j}{\partial t} =&~ \left [ \left ( \bm{\nabla} \psi_0 \right ) \times \left ( \bm{\nabla} + i\mathbf{q}_j^0 \right ) \right ]_z \mathcal{L}_{j} \left ( \alpha_1 + \alpha_2 \mathcal{L}_{j} \right ) A_j \nonumber\\
    &~ + \left \{ \left [ \left ( \bm{\nabla} + i\mathbf{q}_j^0 \right ) A_j \right ] \times \bm{\nabla} \left ( \alpha_1 \nabla^2 \psi_0 + \alpha_2 \nabla^4 \psi_0 \right ) \right \}_z \nonumber\\
    &~ + \sum_{l \neq k \neq j} \left \{ \left [ \left ( \bm{\nabla} - i\mathbf{q}_l^0 \right ) A_l^* \right ] \times \left ( \bm{\nabla} - i\mathbf{q}_k^0 \right ) \right \}_z \nonumber\\
    &\qquad \mathcal{L}_{k}^* \left ( \alpha_1 + \alpha_2 \mathcal{L}_{k}^* \right ) A_k^* - \tilde{q}_0^2 \frac{\delta\mathcal{F}_{\rm PFC}}{\delta A_j^*}, \label{eq:Aj}
\end{align}
where ``$*$'' denotes complex conjugate, $\mathcal{L}_{j} = \nabla^2 + 2i\mathbf{q}_j^0 \cdot \bm{\nabla}  - \tilde{q}_0^2$, and $\mathcal{F}_{\rm PFC}[A_j,\psi_0]$ is the mesoscale effective free energy functional (SI Appendix).

\subsection*{One-Mode Approximation for the Bulk State}

For a pristine crystalline phase of hexagonal symmetry, $\psi_0=\bar{\psi}_0$ and $A_j=A_j^0$ ($j=1,2,3$) are constant in the one-mode approximation. Thus the contributions from transverse interaction terms vanish, as seen from Eqs.~\ref{eq:psi0} and \ref{eq:Aj}, and the exact solution of the steady state is given by
\begin{align}
    A_j^0 &= A_0 = \frac{1}{15} \Big \{ g - 3\bar{\psi}_0  \label{eq:Aj^0}\\
    & + \sqrt{(g-3\bar{\psi}_0)^2 - 15\left [ -\epsilon + (\tilde{q}_0^2 - q_0^2)^2 + 3\bar{\psi}_0^2 - 2g\bar{\psi}_0 \right ]} \Big \}. \nonumber
\end{align}
This agrees with the result of the original PFC model without transverse interactions. The one-mode phase diagram for solid and liquid phases \cite{Provatas10} is the same as well. This is consistent with the microscopic picture of a perfect crystalline bulk state where neighboring transverse forces cancel out and thus have a negligible effect on the system stability, unless there exist elastoplastic deformations or interfaces with varying structural amplitudes, as will be addressed below.

\subsection*{Odd Elasticity}

For small elastic deformations, the phases of complex amplitudes vary at long wavelength as a result of a displacement field $\mathbf{u}$, giving $A_j \simeq A_j^0 \exp(-i\mathbf{q}_j^0 \cdot \mathbf{u})$, where $A_j^0 \simeq A_0$ in the steady state with $\psi_0 \simeq \bar{\psi}_0$. From the amplitude equations \ref{eq:psi0} and \ref{eq:Aj}, we show in SI Appendix that the displacement field $u_{j=x,y}$ satisfies the elastodynamical equation (i.e., the overdamped Cauchy’s equation of motion)
\begin{equation}
    \Gamma \frac{\partial u_j}{\partial t} = \partial_i \sigma_{ij} = C_{ijkl} \partial_i \partial_k u_l,
    \label{eq:uj_elastodynamic}
\end{equation}
where $\Gamma$ is the drag coefficient, $C_{ijkl}$ represents the elastic constant tensor, and $\sigma_{ij} = C_{ijkl} \partial_k u_l$ is the stress tensor. To lowest order,
\begin{align}
    & C_{1111} = C_{2222} = \Gamma \tilde{q}_0^2 \left ( 5\tilde{q}_0^2 - 2q_0^2 \right ), \nonumber\\
    & C_{1122} = C_{2211} = \Gamma \tilde{q}_0^2 \left ( 2q_0^2 - \tilde{q}_0^2 \right ), \nonumber\\
    & C_{1212} = C_{2121} = C_{1221} = C_{2112} = \Gamma \tilde{q}_0^2 \left ( 3\tilde{q}_0^2 - 2q_0^2 \right ),
\end{align}
as in the original PFC model with regular, even elasticity (where $\tilde{q}_0^2 = q_0^2$ and $\Gamma = 3A_0^2$ \cite{ElderPRE10}). The nonzero transverse interactions lead to additional elastic constants $C_{iikl}$ ($k \neq l$) and $C_{ijkk}$ ($i \neq j$), given by
\begin{align}
    & C_{1112} = C_{1121} = -\frac{1}{2}\Gamma A_0 \tilde{q}_0^2 \left (5\alpha_1 - 13\tilde{q}_0^2\alpha_2 \right ), \nonumber\\
    & C_{2221} = C_{2212} = -C_{1112} = -C_{1121}, \nonumber\\
    & C_{1211} = -C_{2122} = \frac{1}{2}\Gamma A_0 \tilde{q}_0^2 \left (7\alpha_1 - 23\tilde{q}_0^2\alpha_2 \right ), \nonumber\\
    & C_{2111} = -C_{1222} = \frac{3}{2}\Gamma A_0 \tilde{q}_0^2 \left (\alpha_1 - \tilde{q}_0^2\alpha_2 \right ), \label{eq:Cijkl}
\end{align}
where the right mirror symmetry $C_{iikl} = C_{iilk}$ is still maintained due to global rotational invariance of the system and the lack of coupling to rotational degrees of freedom, while generally the left minor symmetry is broken, i.e., $C_{ijkk} \neq C_{jikk}$, as a result of nonzero internal torque caused by transverse interactions, which leads to $\sigma_{ij} \neq \sigma_{ji}$ (if $i \neq j$) and the breaking of angular momentum conservation. Notably, the major symmetry is no longer obeyed, with $C_{ijkk} \neq C_{kkij}$ and $C_{iikl} \neq C_{klii}$ in Eq.~\ref{eq:Cijkl}, such that the system incorporates both even and odd elasticity, i.e.,
\begin{equation}
    C_{iikl} = C_{iikl}^{\rm (e)} + C_{iikl}^{\rm (o)}, \qquad
    C_{ijkk} = C_{ijkk}^{\rm (e)} + C_{ijkk}^{\rm (o)},
\end{equation}
consisting of even elasticity components $C^{(\rm e)}_{ijkl} = C^{(\rm e)}_{klij}$ satisfying major symmetry and odd elasticity components $C^{(\rm o)}_{ijkl} = - C^{(\rm o)}_{klij}$ which are antisymmetric and proportional to transverse interaction strength (with expressions determined by Eq.~\ref{eq:Cijkl} and shown in SI Appendix). The bulk modulus $B$ and shear modulus $\mu$ are then given by
\begin{align}
    B &= \frac{1}{2} \left ( C_{1111} + C_{1122} \right ) = 2\Gamma \tilde{q}_0^4, \nonumber\\
    \mu &= \frac{1}{2} \left ( C_{1111} - C_{1122} \right ) = \Gamma \tilde{q}_0^2 \left ( 3\tilde{q}_0^2 - 2q_0^2 \right ).
\end{align}
We can also identify the odd bulk modulus $A$ induced by nonzero internal torque and the odd shear modulus $K^{\rm o}$ as
\begin{align}
    A &= \frac{1}{2} \left ( C_{2111} - C_{1211} \right ) = -\Gamma A_0 \tilde{q}_0^2 \left (\alpha_1 - 5\tilde{q}_0^2\alpha_2 \right ), \nonumber\\
    K^{\rm o} &= C_{1112} = -\frac{1}{2} \left (C_{2111} + C_{1211} \right ) \nonumber\\
    & = -\frac{\Gamma}{2} A_0 \tilde{q}_0^2 \left (5\alpha_1 - 13\tilde{q}_0^2\alpha_2 \right ).
\end{align}
All these are consistent with continuum odd elasticity theory \cite{ScheibnerNatPhys20}, with elastic constants expressed explicitly in terms of T-PFC parameters.

\section*{Results and Discussion}

\subsection*{Model Validation}

\begin{figure*}
\centering
\includegraphics[width=\linewidth]{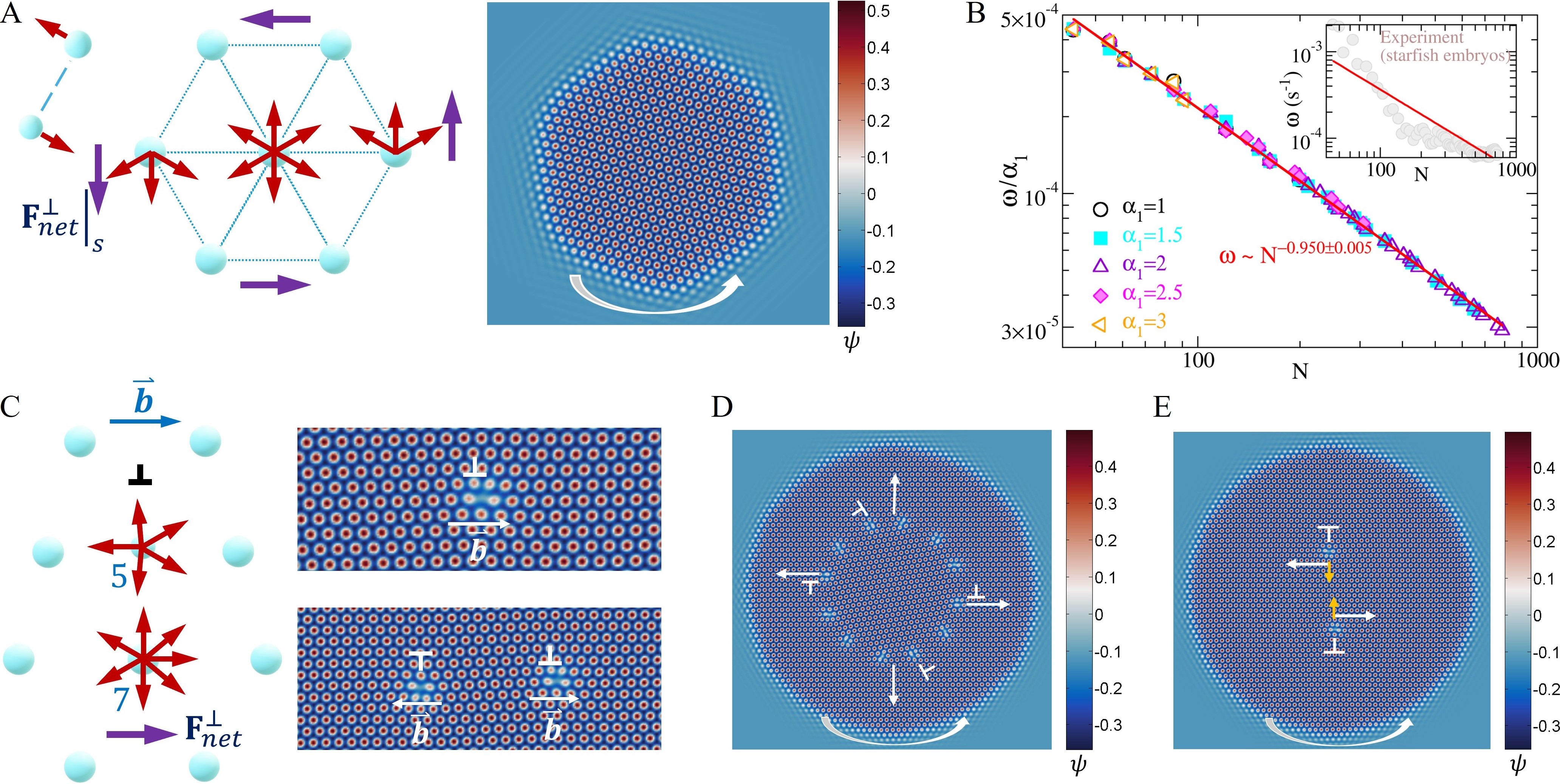}
\caption{T-PFC simulation results for self-rotating odd crystals and motile defects. (A) A snapshot of a self-rotating single-crystalline grain at $\alpha_1=2$, with a schematic illustrating nonzero net transverse force on the surface. (B) The scaling of grain rotation frequency $\omega$ vs particle number $N$, showing a data collapse for different transverse interaction strength $\alpha_1$. The inset shows the experimental data of starfish-embryos living crystals in Ref.~\cite{TanNature22} for the same $N$ range. The simulation results are fitted into a power law scaling which is also indicated in the inset for comparison. (C) Schematic of a penta-hepta dislocation with local nonzero net transverse force, which leads to dislocation self-glide and the unbinding of a dislocation dipole as shown in the snapshots of the simulated $\psi$ profile. The Burgers vector $\mathbf{b}$ is labeled for each dislocation. (D) A self-expanding circular grain boundary within a self-rotating crystallite, with $8^\circ$ misorientation at $\alpha_1=1$. (E) The climb and self-glide of two dislocations of opposite Burgers vectors at $\alpha_1=1$. Small arrows indicate the climb direction, and large arrows give the direction of self-glide along the Burgers vectors.}
\label{fig:model_val}
\end{figure*}

In addition to the above analytic results which demonstrate that the T-PFC model developed here incorporates key aspects of the system symmetry (and symmetry breaking), chirality, and odd elastic properties, we have conducted numerical simulations to further validate the model, and to provide evidence for the multiscale nature of this approach where particle-scale resolution is still maintained in this continuum field description. Here the T-PFC model equation \ref{eq:TPFC} is solved numerically (Materials and Methods; see also SI Appendix), with the setting of $\alpha_2=0$ for a minimal continuum field model incorporating effects of transverse interaction and odd elasticity. (We have also conducted simulations with nonzero $\alpha_2$ and found no significant differences.) Values of model parameters used in our simulations are summarized in SI Appendix, Table S1. In SI Appendix, we also present a procedure of model parameterization to match the model to experimental systems. We have reproduced various phenomena of odd and chiral crystals found in experiments and particle-based simulations and matched the results to the microscopic description based on interparticle forces, with some simulation outcomes illustrated in Fig.~\ref{fig:model_val}. These include self-rotation of odd crystallites (Fig.~\ref{fig:model_val} A and B and Movie S1), as a result of nonzero net transverse force (or net odd stress) and thus a net torque on the free surface of a crystallite embedded in a homogeneous or liquid-state medium, consistent with experiments of, e.g., swimming bacteria \cite{PetroffPRL15}, starfish embryos \cite{TanNature22}, and magnetic colloids \cite{BililignNatPhys22} 2D chiral crystals. Also included are the behaviors of dislocation self-propulsion (Fig.~\ref{fig:model_val}C). A key phenomenon is the self-glide of a dislocation along the direction of its Burgers vector (Movie S2), which follows the direction of local net transverse force around the dislocation core as indicated in the schematic of Fig.~\ref{fig:model_val}C. For a dislocation dipole with opposite Burgers vectors and far enough separation, dislocation unbinding occurs (Movie S3), instead of annihilation in regular crystals (Movie S4). A circular grain boundary would either expand (Fig.~\ref{fig:model_val}D and Movie S5) or shrink (Movie S6), depending on the direction of Burgers vectors of boundary dislocations. These results of motile dislocation are consistent with those found in magnetic colloidal experiments and particle-based simulations \cite{BililignNatPhys22,BravermanPRL21,PoncetPRL22}, demonstrating that effects at microscopic, lattice scales (that are not accessible in standard continuum approaches) can be well captured in this continuum T-PFC model.

An advantage of PFC modeling is that it allows to simulate the process of dislocation climb, which is difficult to realize in atomistic simulations like MD as the dynamics involves long-range mass diffusion. As shown in Fig.~\ref{fig:model_val}E, two dislocations of opposite sign of Burgers vectors, which would climb towards each other and annihilate in regular crystals (Movie S7), travel apart due to self-gliding caused by transverse interaction (Movie S8). Note that in all the calculations presented here, we choose $\alpha_1>0$ which corresponds to the case of two-body transverse force along the direction $\hat{\mathbf r}_{ab}^\perp = \hat{\mathbf z} \times \hat{\mathbf r}_{ab}$. This leads to counterclockwise self-rotation of crystallites and the dislocation self-propulsion along the direction of Burgers vector, while choosing $\alpha_1<0$ with opposite direction of interparticle transverse force gives equivalent results, although with opposite direction of self-rotation (clockwise) and dislocation self-glide, as verified in our simulations.

We also quantitatively examine the property of self-rotation of single-crystalline grains, for which the grain rotates faster for larger transverse interaction strength $\alpha_1$ and smaller grain size, consistent with the microscopic picture. Our T-PFC calculations of grain self-rotation frequency $\omega$ at different $\alpha_1$ show the collapse of data into a power law scaling (Fig.~\ref{fig:model_val}B)
\begin{equation}
    \frac{\omega}{\alpha_1} \sim N^{-s}, \label{eq:omega_N}
\end{equation}
with scaling exponent $s = 0.950 \pm 0.005$, where $N$ is the number of particles in the grain. For comparison, we show in the inset of Fig.~\ref{fig:model_val}B the data of the experimental measurement of rotating starfish embryo clusters \cite{TanNature22} within the same range of $N$. Although the experimental data indicates a crossover between two regimes of $\omega$ scaling for small and large embryo number $N$, which might involve some biological mechanisms and could not be captured by our T-PFC continuum field modeling and by the related particle-based simulation \cite{TanNature22}, the overall trend of the $\omega$ variation is consistent with the scaling behavior identified in our modeling. To understand this $N^{-1}$ scaling, consider the overdamped limit with $\zeta_R \omega = \tau_{\rm net}$, where $\zeta_R$ is the rotational friction coefficient that can be estimated as $\zeta_R \propto R^4 \sim N^2$ for a 2D circular grain of radius $R$ (SI Appendix), and the self-generated net torque on the grain surface $\tau_{\rm net} = \int_s R\, dF^\perp_{\rm net} \propto \oint R \alpha_1 ds \propto \alpha_1 R^2$ (if assuming balanced transverse interactions in the bulk). This leads to $\omega = \tau_{\rm net} / \zeta_R \propto \alpha_1/R^2 \sim \alpha_1/N$.

\subsection*{Surface Cusp Instability and Grain Self-Fission}

\begin{figure}
    \centering
    \includegraphics[width=0.95\linewidth]{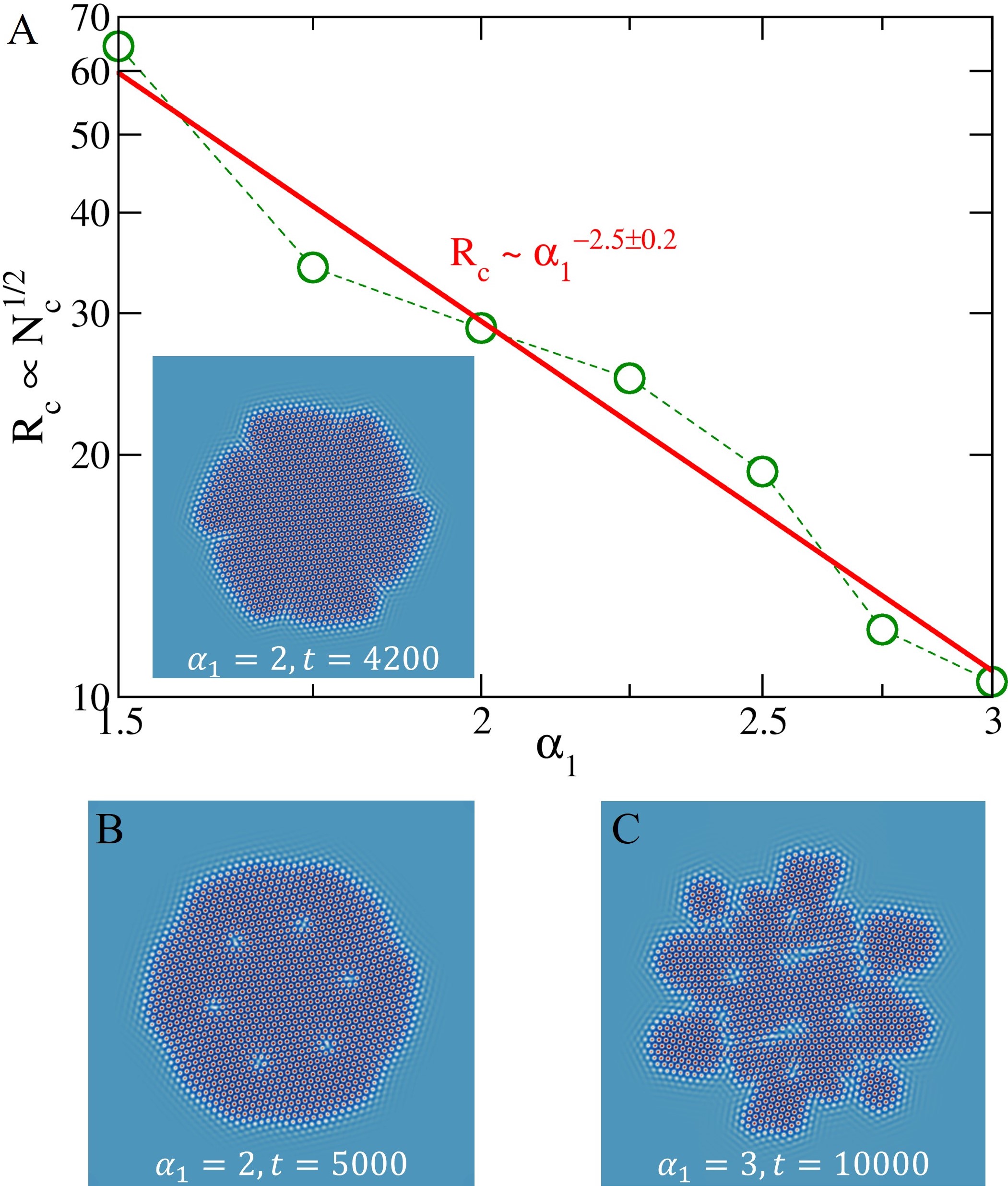}
    \caption{(A) Critical grain radius $R_c$ for the onset of surface cusp instability as a function of transverse interaction strength $\alpha_1$, as identified from T-PFC simulations (up to $t=10^6$). Snapshots of the $\psi$ profile are shown in the inset of (A) for the instability onset and in (B) for the surface-emitted motile dislocations at $\alpha_1=2$, and in (C) for grain self-fission at $\alpha_1=3$.}
    \label{fig:surf_inst}
\end{figure}

On the free surface of single-crystalline odd grains, the self-generated surface odd stress (with tangential, self-shearing surface net transverse force as illustrated in Fig.~\ref{fig:model_val}A) results in the spontaneous development of a surface cusp instability in the absence of any external strain or stress. An example with $\alpha_1=2$ is shown in the inset of Fig.~\ref{fig:surf_inst}A and Movie S9. The subsequent nonlinear evolution leads to the nucleation of dislocations at the edges of surface cusps, which self-propel into the bulk as a result of the self-gliding of motile dislocation induced by transverse interaction (Fig.~\ref{fig:surf_inst}B). The instability occurs beyond a threshold of grain size, characterized by a critical radius $R_c$ which is larger for weaker transverse interaction strength $\alpha_1$. A scaling behavior of $R_c \sim |\alpha_1|^{-\beta}$ is identified from our numerical simulations (Fig.~\ref{fig:surf_inst}A), with a scaling exponent $\beta=2.5 \pm 0.2$.

Note that the odd crystalline system studied here is nonpotential as a result of nonzero transverse interactions, and the system evolution governed by Eq.~\ref{eq:TPFC} does not follow the pathway of free energy minimization. Thus this result of grain instability is of nonequilibrium, nonrelaxational nature and cannot be identified from energetics arguments used for near-equilibrium potential systems (see SI Appendix for a related calculation). The physical mechanism for this behavior of critical radius and scaling is governed by the competition between the destabilizing effect caused by the self-generated surface odd stress and the stabilizing effect of surface tension, and can be understood through the mechanical force/stress condition on the surface. At a crystal-melt interface $\Sigma$, the mechanical force balance is given by \cite{LeoActaMetall89}
\begin{equation}
    \bm{\sigma} \cdot \hat{\mathbf n} + P_L \hat{\mathbf n} - \bm{\nabla}_{\Sigma} \cdot \bm{\sigma}_s = 0,
    \label{eq:mech_condition}
\end{equation}
where $\hat{\mathbf n}$ is the surface unit normal, $\bm{\sigma}$ is the stress tensor of the crystallite, $P_L$ is the pressure of the surrounding liquid or melt, $\bm{\nabla}_{\Sigma}$ represents the surface gradient, and $\bm{\sigma}_s$ is the surface stress tensor. For a steady-state 2D circular crystallite with radius $R$, $\hat{\mathbf n} = \hat{\mathbf r}$ with unit vector $\hat{\mathbf r}$ along the radial direction, $\bm{\nabla}_{\Sigma} \cdot \bm{\sigma}_s = -\sigma_s \kappa \hat{\mathbf n} = -(\sigma_s/R) \hat{\mathbf r}$ with surface curvature $\kappa=1/R$, and $\bm{\sigma} \cdot \hat{\mathbf n}|_R = -P_S \hat{\mathbf r}$ for hydrostatic elastic bulk deformations, with no contribution from the net odd stress on the surface along the radial direction (see Fig.~\ref{fig:model_val}A). We then have
\begin{equation}
    P_S - P_L - \frac{\sigma_s}{R} = 0,
\end{equation}
which is the 2D solid-state analog of the Young-Laplace equation, but with surface energy replaced by surface stress $\sigma_s$ \cite{CahnActaMetall80,MullinsJCP84,LeoActaMetall89}. Now consider a small surface perturbation or deformation that leads to surface normal deviation from $\hat{\mathbf r}$, i.e., $\hat{\mathbf n} \rightarrow \hat{\mathbf r} + \delta\hat{\mathbf n}$. The odd stress $\bm{\sigma}^{(\rm o)}$ should then give a nonzero contribution along the local surface normal at $r \approx R$, with $\bm{\sigma}^{(\rm o)} \cdot \hat{\mathbf n} |_R \simeq \delta\sigma_R^{(\rm o)} \delta\hat{\mathbf n}$, where the odd stress component on the surface, $\delta\sigma_R^{(\rm o)}$, is proportional to transverse interaction strength $\alpha_1$ and should weakly increase with smaller $R$ due to a relatively stronger effect of deformation for smaller grain with a perturbed surface. This weak variation of $\delta\sigma_R^{(\rm o)}$ would thus follow a sublinear decay over $R$, and we can expand in terms of a fractional power series (i.e., Puiseux series \cite{Puiseux1850}) of $1/R$ with a small leading-order fractional exponent to account for this sublinear variation. Noting that $R$ needs to be large enough for a nucleated solid grain to avoid being dissolved, it is hence a good approximation to keep the leading-order term of expansion which yields $\delta\sigma_R^{(\rm o)} \simeq |\alpha_1| c_T (1/R)^{\delta}$, with a fractional exponent $0 < \delta <1$ and a proportional constant $c_T$. The surface perturbation also leads to a modification of surface stress contribution, with $\bm{\nabla}_{\Sigma} \cdot \bm{\sigma}_s \simeq -\sigma_s \kappa \hat{\mathbf n} - \delta\sigma_s \kappa \delta\hat{\mathbf{n}}$. Making use of the mechanical force condition (Eq.~\ref{eq:mech_condition}), we can then determine the surface perturbed velocity
\begin{equation}
    \mathbf{v}_s \propto \left ( \delta\sigma_R^{(\rm o)} - \delta\sigma_s \kappa \right ) \delta\hat{\mathbf{n}} \simeq \bigg( |\alpha_1| c_T R^{-\delta} - \frac{\delta\sigma_s}{R} \bigg) \delta\hat{\mathbf{n}}.
\end{equation}
The odd crystallite surface instability would occur when $\delta\sigma_R^{(\rm o)} > \delta\sigma_s \kappa$ at $R > R_c$, resulting in a critical radius
\begin{equation}
    R_c \propto |\alpha_1|^{-1/(1-\delta)},
\end{equation}
and thus a scaling exponent $\beta = 1/(1-\delta)$. The above simulation result gives an exponent of $\beta \simeq 5/2$, indicating $\delta \simeq 3/5$. This is consistent with the assumption of small fractional $\delta$ for a sublinear variation of odd stress contribution on the perturbed surface.

At the later time stage of grain evolution, the continuing process of dislocation proliferation and self-traveling results in a recurring procedure of cracking and self-healing inside the rotating crystallite, yielding internal grain boundary formation and evolution as well as the overall grain distortion (see Movie S9 at $\alpha_1=2$). In the case of strong enough transverse interaction, it leads to the occurrence of grain self-fission with self-rotating fragments (Fig.~\ref{fig:surf_inst}C and Movie S10 at $\alpha_1=3$). Recent MD simulations and experiments of magnetic colloidal crystals \cite{BililignNatPhys22} have observed the similar phenomenon of dislocation creation and propulsion from the free boundary of single-crystalline crystallite and the subsequent formation of multidomains and grain boundaries inside the crystallite, although the mechanism of cusp instability and its grain size dependence, as well as the self-division and the resulting grain distortion as found here, were not examined. Thus the mechanism and results identified above are key to understanding those experimental findings. Note also that the phenomena of cusp instability and irregular grain self-fission obtained here for ``dry'' odd crystals originate from self-generated surface odd stresses resulting from transverse interactions, a mechanism that is intrinsically different from the self-shearing instability and flow-induced splitting of fluid droplets found in the ``wet'' system of Active Model H \cite{SinghC2019}. Such a ``wet'' active model system is subjected to hydrodynamic flow coupling but not to transverse interactions (i.e., without any odd elastic or viscous properties), where the (not crystalline) fluid droplets form via phase separation and have negative surface tension caused by contractile active stress, leading to a self-shearing instability and droplet breaking. This mechanism of instability then differs fundamentally from the one described above for odd crystals, due to the lack of odd elasticity. The critical droplet size for the onset of instability and the corresponding scaling behavior, which have not been examined for this ``wet'' fluid system, are then expected to be qualitatively different as well.

\subsection*{Transition from Normal to Reverse Ostwald Ripening}

\begin{figure}[tb]
\centerline{\includegraphics[width=0.95\linewidth]{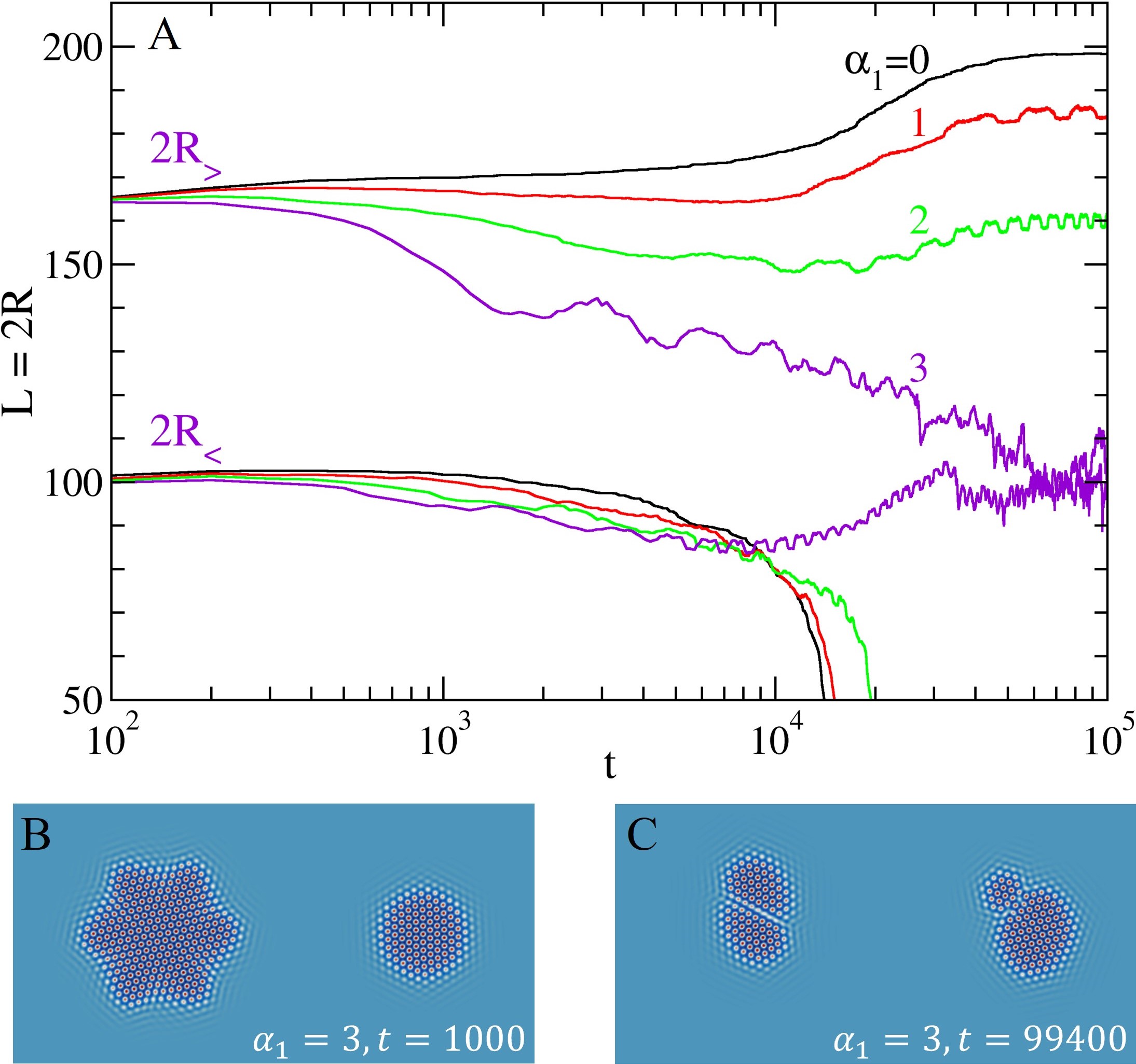}}
\caption{(A) Time evolution of grain sizes during the ripening process of two crystalline grains, showing a transition from normal to reverse Ostwald ripening with the increase of $\alpha_1$. (B and C) Simulation snapshots at $t=1000$ and $99400$ during the process of reverse Ostwald ripening with $\alpha_1=3$.}
\label{fig:Ostwald_ripening}
\end{figure}

This mechanism of self-stress induced surface instability and the resulting grain self-fission can lead to two phenomena of anomalous multigrain dynamics. The first one is the transition from normal Ostwald ripening to the reverse process with the increase of transverse interaction strength $\alpha_1$, for two self-rotating odd crystalline grains. This is demonstrated in Fig.~\ref{fig:Ostwald_ripening}A, which shows the time evolution of grain sizes, defined as $L \simeq 2R=2\sqrt{{\rm \mathcal{A}}/\pi}$ with grain area $\mathcal{A}$, at $\alpha_1$ ranging from $0$ (for the original passive PFC model) to $3$. The standard process of Ostwald ripening is observed at small enough magnitude of $\alpha_1$ with zero or weak enough transverse interactions, as expected, for which the initially larger grain with effective radius $R_>(t)$ grows with time after an early time range of relaxation, accompanied by the shrinkage of the initially smaller grain with radius $R_<(t)$. At strong enough $\alpha_1$, starting from the same initial conditions of grain sizes and spacing, reverse Ostwald ripening occurs when the initial radius $R_<(0)$ of the smaller grain is large enough, showing as the overall decrease of $R_>(t)$ with time and the early-time decrease, slowing-down, and then increase of $R_<(t)$ until both reach a similar scale (Fig.~\ref{fig:Ostwald_ripening} B and C; see also Movie S11 (for $t \leq 20000$) and Movie S12 (for $t=95000$--$10^5$), corresponding to the curves of $\alpha_1=3$ in Fig.~\ref{fig:Ostwald_ripening}A). 

This transition can be explained by the behavior of grain self-division and fission. In the case of normal Ostwald ripening, according to the Gibbs-Thomson relation \cite{LangerRMP80} (which has been generalized for the PFC model with the supersaturation of density variation field $\psi$ \cite{HuangPRE13}), i.e., $\delta\mu|_{\rm surface} \propto \gamma\kappa$ with $\delta\mu = \mu - \mu_{\rm eq}$, $\gamma$ the surface tension, and $\kappa$ the surface curvature, the chemical potential $\mu$ and thus particle density at the solid-liquid grain surface increase with the decreasing grain radius (i.e., larger $\kappa$), leading to a diffusive mass transport from smaller to larger grains as driven by the local density gradient and consequently, the growth of larger grain at the expense of shrinking and disappearance of smaller grain. This can be changed by strong enough transverse interactions. For a given initial radius $R_>$ of the larger self-rotating grain, when it exceeds the threshold size $R_c(\alpha_1)$, at a high enough $\alpha_1$ (as quantified in Fig.~\ref{fig:surf_inst}A) a surface cusp instability and the subsequent dislocation proliferation, emission and grain self-fission occur, resulting in smaller fragmented sub-grains and thus the reduced or even reversed local density gradient and diffusion flux with respect to the other grain (which has size $R_<$). This leads to the slowing-down, arrest, and reverse of the Ostwald ripening process with the two grains approaching a similar, comparable size, as shown in Fig.~\ref{fig:Ostwald_ripening} (for results of $\alpha_1=3$). At the late time stage when the radius of the initially smaller grain $R_<(t) > R_c$, grain self-fission also occurs there, in addition to the self-division of the other grain (Fig.~\ref{fig:Ostwald_ripening}C and Movie S12). On the other hand, if the initial $R_<(0)$ is too small, due to the Gibbs-Thomson effect the smaller grain shrinks rapidly and melts before the self-fission and fragmentation of the larger grain take any effect. 

This mechanism governing the reverse Ostwald ripening process for odd crystals is thus different from that of phase-separated active fluids caused by negative pseudotension of liquid droplet surface \cite{TjhungNC2018,SinghC2019}, noting the increase of positive $\gamma$ with stronger transverse interactions (which generate larger tangential surface stress) as found in simulations of chiral fluid interfaces \cite{CaporussoPRL24}. It is also different from the reverse ripening caused by misfit stress in passive solid systems \cite{SuActaMater96}, given the lack of any imposed stress or strain in the system studied here. Hence the transition identified above is a unique feature combining chirality (breaking of 2D parity) and grain self-fission in this odd crystalline system.

\subsection*{Transition from Grain Coarsening to Grain Self-Fragmentation}

\begin{figure*}[tb]
\centering
\includegraphics[width=\linewidth]{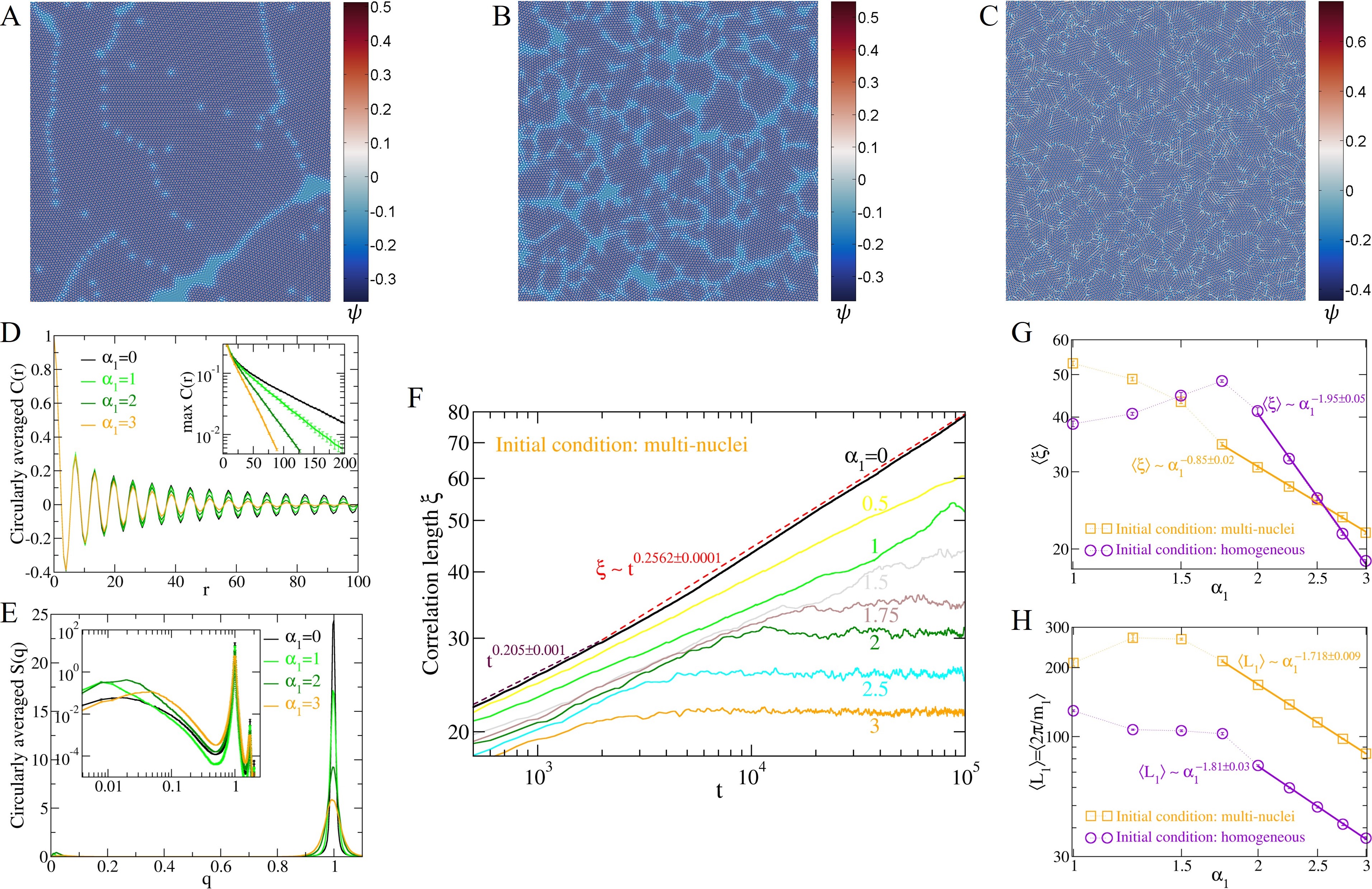}
\caption{Multigrain dynamics of the odd polycrystalline state. (A-C) Simulation snapshots at $t=10^5$, each showing the central quarter of the $2048 \times 2048$ simulated system, for initial conditions of multiple crystalline nuclei with $\bar{\psi}_0=-0.09$ at (A) $\alpha_1=1$ governed by grain coarsening dynamics and (B) $\alpha_1=3$ governed by grain self-fragmentation, and (C) from homogeneous initial state with $\bar{\psi}_0=0$ at $\alpha_1=3$. (D) Circularly averaged correlation function $C(r)$, with inset showing the exponential decay of its peak values, and (E) circularly averaged structure factor $S(q)$, with the corresponding log-log plots given in the inset, both at $t=10^5$ and for initial conditions of multiple nuclei. (F) Time variation of the correlation length $\xi$, showing a transition from grain coarsening to self-fragmentation dynamics with the increase of transverse interaction strength $\alpha_1$. (G and H) $\langle \xi \rangle$ and $\langle L_1 \rangle$, averaged over the late time stage ($t=8 \times 10^4$--$10^5$), as a function of $\alpha_1$, for two types of initial conditions. Power-law scalings are identified in the regime of grain self-fragmentation. Results in (D-H) have been averaged over 20 simulation runs for $2048 \times 2048$ system size.}
\label{fig:multigrains}
\end{figure*}

The second phenomenon originating from the mechanism of surface cusp instability and grain self-fission is the spontaneous formation of a dynamical polycrystalline state during the nucleated growth process, where the system evolves from an initial condition of multiple randomly distributed crystalline nuclei. Examples of system time evolution at small and large $\alpha_1$ are shown in Movies S13 ($\alpha_1=1$) and S14 ($\alpha_1=3$), with some simulation snapshots given in Fig.~\ref{fig:multigrains}, both having the same average density $\bar{\psi}_0=-0.09$ and starting from the same initial conditions. At the early time stage, a similar behavior is observed in both movies, including the fast growth of nuclei and the impingement and then coalescence of grains, but the subsequence evolution is qualitatively different. At small $\alpha_1$ (weak transverse interactions), the system dynamics is governed by the motion of topological defects, particularly grain boundaries, similar to the typical process of grain coarsening, although with dislocation self-motion as a result of nonzero transverse force (Fig.~\ref{fig:multigrains}A and Movie S13). However, for large $\alpha_1$ (strong transverse interaction), any large enough self-rotating grains, formed after either grain growth or coalescence, exhibit self-fission into smaller parts through dislocation creation on grain surface and propagation across the grain with crack formation. Neighboring rotating grains then remerge or reheal, self-splitting again if exceeding the instability size threshold ($R_c$), with the procedure iterated incessantly. This leads to a persistently varying process of grain self-fragmentation instead of coarsening, as seen in Movie S14 and Fig.~\ref{fig:multigrains}B. 

To quantitatively examine this transition from the normal grain coarsening process to the anomalous dynamics of grain fragmentation, we calculate the structure factor $S(\mathbf{q},t) = |\hat{\psi}_\mathbf{q}(t)|^2/V$, with $\hat{\psi}_\mathbf{q}(t)$ the Fourier transform of $\psi(\mathbf{r},t)$ and $V=L_xL_y$ for a 2D system of size $L_x \times L_y$, and the correlation function $C(\mathbf{r},t)=(\langle \psi(\mathbf{x}+\mathbf{r},t) \psi(\mathbf{x},t) \rangle - \bar{\psi}_0^2) / (\langle \psi^2 \rangle - \bar{\psi}_0^2)$ at different times $t$. The characteristic domain or grain size in the polycrystalline state is identified by two quantities, the correlation length $\xi(t)$ determined via the fitting of the envelope (or peak values) of circularly averaged correlation function to $C_m(r,t) \propto \exp(-r/\xi)$ (Fig.~\ref{fig:multigrains}D), and domain size $L_1(t) = 2\pi/m_1$ where the first moment $m_1 = \int_0^{q_{m_1}} dq\,q S(q,t) / \int_0^{q_{m_1}} dq S(q,t)$ is evaluated within the first peak region of small wave numbers $0 < q \leq q_{m_1}$, with $q_{m_1}$ the location of first local minimum of the circularly averaged structure factor $S(q,t)$ (see Fig.~\ref{fig:multigrains}E, where the other peaks are located at the wave numbers of triangular lattice with ratios $1:\sqrt{3}:2$). Note that $\xi$ measures the distance of correlation between local density fluctuations, giving the scale of domains separated by any defects or disorder, while $L_1$ characterizes the larger-scale profile of solid grains particularly those separated via locally homogeneous or liquid-like boundaries or gaps.

Results of time evolution of the correlation length $\xi$ are presented in Fig.~\ref{fig:multigrains}F, clearly showing a transition from coarsening-dominated to fragmentation-dominated dynamics with the increase of transverse interaction strength $\alpha_1$. A similar behavior is obtained for $L_1$ from calculations of the first moment. The self-fragmentation state appears earlier for stronger transverse interaction, as characterized by a fluctuating plateau of the average grain size which decreases with larger $\alpha_1$ as a result of more dynamically fragmented grains. This is consistent with the $\alpha_1$-dependence of threshold $R_c$ for grain instability found in Fig.~\ref{fig:surf_inst}A, although with different values of the scaling exponent. The scalings of characteristic domain sizes averaged over a late time stage (e.g., $t=8 \times 10^4$--$10^5$) are shown in Fig.~\ref{fig:multigrains} G and H, giving $\langle \xi \rangle \sim |\alpha_1|^{-\beta_\xi}$ and $\langle L_1 \rangle \sim |\alpha_1|^{-\beta_{m_1}}$ in the fragmentation-dominated regime, with scaling exponents $\beta_\xi = 0.85 \pm 0.02$ and $\beta_{m_1} = 1.718 \pm 0.009$ respectively (similar results are obtained if choosing different late time ranges for averaging or different system sizes). The transition point between regimes of two different grain dynamics can be estimated as the beginning of the scaling regime, which occurs around $\alpha_1 \simeq 1.75$ as identified from Fig.~\ref{fig:multigrains} G and H.

This phenomenon of multigrain self-fragmentation dynamics closely resembles the self-kneading polycrystal whorl state observed in recent experiments of magnetic colloidal odd crystals \cite{BililignNatPhys22}, where the average grain size is also found to decrease with stronger transverse interaction at higher colloid spinning frequency. This phenomenon, however, is fundamentally different from the clustering or rotating flocks observed in active chiral fluids composed of spinning colloidal particles \cite{LiebchenPRL17,ZhangNatCommun20,Massana-CidPRR21,CaporussoPRL24}. There, although a steady state of finite size clusters without coarsening can form at high enough spinning rates, each of self-rotating clusters or flocks is in the liquid phase with the presence of edge current \cite{Massana-CidPRR21,CaporussoPRL24} but without the mechanism of surface instability or grain self-fission found in the odd crystals studied here. Note that although the dynamical regime of grain fragmentation observed here involves the self-motion and flow or propagation of defects, intrinsically these defect self-flows still originate from the effects of transverse interactions and the associated odd stresses, while effects of hydrodynamic flow coupling by the liquid or melt surrounding the fragmented grains are not considered in this modeling and thus do not contribute to the properties identified above.

In the above study of multigrain dynamics, the average density of the system ($\bar{\psi}_0=-0.09$) is chosen such that some small liquid-state regions or gaps can coexist with solid grains, facilitating the occurrence of grain surface instability. Our simulations indicate that the grain boundaries themselves, without any surrounding liquid, can also develop an instability and emit motile dislocations, causing self-fission of grains. We thus simulate systems of pure solid phase with high enough average density ($\bar{\psi}_0=0$), starting from initial conditions of spatially homogeneous phase. A polycrystalline state emerges spontaneously, with differently oriented domains separated by grain boundaries and dislocations. Similarly, the subsequent evolution is governed by two distinct types of dynamics, grain coarsening at weak transverse interaction strength $\alpha_1$, and as $\alpha_1$ increases, a transition to the persistent dynamical state of grain self-fragmentation (Movie S15 and Fig.~\ref{fig:multigrains}C), although with a different morphology of self-rotating fragmented multigrains. Such a transition occurs at the similar magnitude of $\alpha_1$, around $\alpha_1 \simeq 2$ as obtained from Fig.~\ref{fig:multigrains} G and H. It also features the power-law scalings of $\langle \xi \rangle$ and $\langle L_1 \rangle$, but with much larger exponents $\beta_\xi = 1.95 \pm 0.05$ and $\beta_{m_1} = 1.81 \pm 0.03$, indicating a more prominent effect of transverse interaction at high strength for the self-breaking of grains from grain boundaries.

We remark that the transition from grain coarsening to fragmentation is associated with a percolating network structure of grain boundaries, with an example shown in Fig.~\ref{fig:multigrains}C. In contrast to defect networks in traditional material systems, the odd defect network emerging here becomes highly dynamical when transitioned to the grain fragmentation state, with network defect structures varying perpetually through the breaking of grain boundaries and grains self-fission and the subsequent remerging. This leads to a persistent process of breaking and restoring of the defect network connectivity, as seen in Movie S15 for a portion of the simulated system.

All the above results of transition from grain coarsening to self-fragmentation are robust as long as the system size is large enough such that individual grains or domains are able to exceed the size threshold to develop instability. At small $\alpha_1$ for which the critical size $R_c$ of grain instability could even reach or exceed the system size, the timescale for the development of instability is much longer than that of the coarsening process which would then dominate. Thus, the transition between the two regimes of multigrain dynamics is expected to be size independent for large systems, as verified in our simulations. We have simulated a larger system of size $4096 \times 4096$ (with 5 independent runs) for both types of initial conditions, and the results obtained, including the transition value of $\alpha_1$ and scaling properties of characteristic domain sizes, are quantitatively very similar to those presented in Fig.~\ref{fig:multigrains} for a $2048 \times 2048$ system (with 20 runs).

\subsection*{Grain Locomotion}

\begin{figure*}[tb]
\centering
\includegraphics[width=0.9\linewidth]{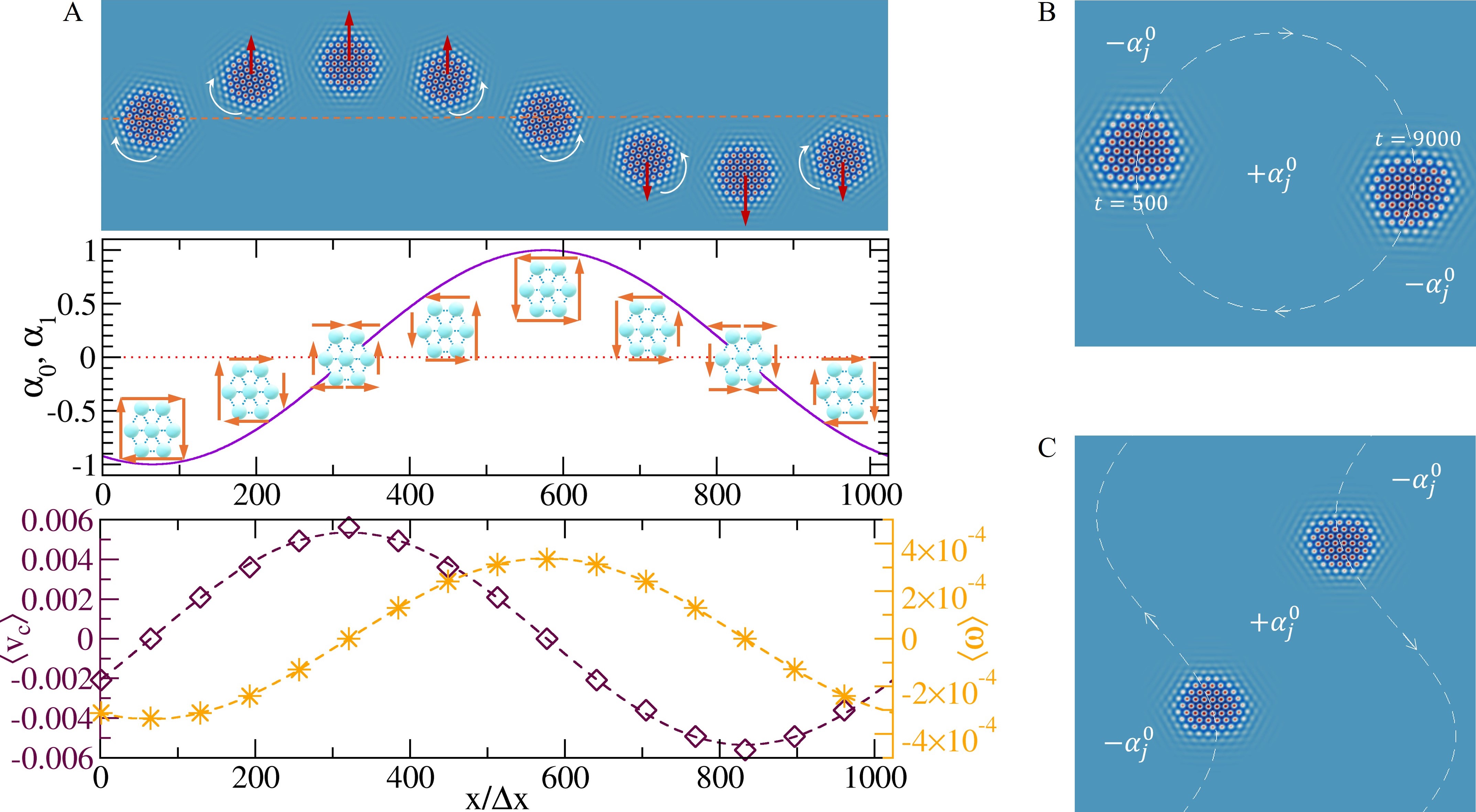}
\caption{Odd grain locomotion simulated via the T-PFC model. (A) A transition from self-rotation, self-rolling, to self-translation of grain locomotion as controlled via spatially-varying transverse interaction strength $\alpha_0=\alpha_1$. The rotational and translational directions of the grain are indicated as arrows in the top panel which shows the simulation snapshot. The arrows in the middle panel illustrate the net surface transverse forces for each grain, and the spatial dependence of self-translation velocity $\langle v_c \rangle$ and self-rotation frequency $\langle \omega \rangle$ (averaged over $t=3000$--$5000$) is given in the bottom panel, with dashed curves being the fitting to the data calculated from simulations. (B and C) The controlled transport of odd elastic grains along (B) a circular trajectory and (C) S-shaped tracks, through the pre-designed 2D spatial distribution of $\alpha_{j=0,1}$.}
\label{fig:locomotion}
\end{figure*}

The surface effect induced by net transverse interaction can be utilized to control the locomotion of individual odd crystalline grains, showing as migration or transport of grains with directional self-motion (either translation or rotation) even though the constituent particles are not self-propelled. This is achieved by imposing spatially varying transverse interaction strengths $\alpha_{j=0,1,2}$ in the T-PFC equation \ref{eq:T-PFC}, which could correspond to spatial variation of individual particle spinning rates (as tuned by, e.g., rotating magnetic field applied to magnetic colloids, or similar avenues) that determine the magnitude of interparticle transverse force. It can be illustrated via an example of one-dimensional variation
\begin{equation}
    \alpha_j(x) = \alpha_j^0 \cos Q_x (x-x_0),
    \label{eq:alphaj_1D}
\end{equation}
where $j=0,1$ and $Q_x=2\pi/\lambda_x$ with $\lambda_x$ the spatial periodicity of $\alpha_j$ variation which can be controlled by, e.g., spatial dependence of magnetic field for spinning magnetic particles. This leads to a continuous transition across different modes of grain locomotion, as demonstrated in Fig.~\ref{fig:locomotion}A and Movie S16 (where $\alpha_0=\alpha_1$ and $\alpha_2=0$). When located at the minimum (with $\alpha_{j=0,1}<0$) and maximum (with $\alpha_{j=0,1}>0$) of the spatially varying $\alpha_j(x)$, the grain self-rotates clockwise (CW) and counterclockwise (CCW) respectively, with the maximum magnitude of rotation frequency $\omega$ but with the absence of translational motion. In contrast, when the grain is centered at $\alpha_j=0$ with antisymmetric distribution of $\alpha_{j=0,1}$ in its two sides, pure self-translation occurs, without any grain rotation, and the grain self-travels along a path perpendicular to the direction of $\alpha_j$ variation. In between these two limits of modes, the grain locomotes via the combination of self-rotating and self-translating, i.e., a behavior of self-rolling, and the directions of motion (CW vs CCW rotation and $+y$ vs $-y$ translation) depends on the type of asymmetric distribution of $\alpha_{j=0,1}$ around the grain.

This transition between different locomotion modes can be understood via the microscopic picture of the corresponding surface transverse forces, as illustrated by the schematics in the middle panel of Fig.~\ref{fig:locomotion}A. Note that nonzero net surface forces (or net surface odd stresses) are generated by the interparticle transverse interaction, as shown in the schematic of Fig.~\ref{fig:model_val}A. If the $x$-varying $\alpha_j$ distribution of transverse interaction strength is symmetric about the grain center, like the one located at the minimum of $\alpha_{j=0,1}$ (the leftmost schematic in the mid panel of Fig.~\ref{fig:locomotion}A), the net surface tangential forces generated at the opposite $x$-sides and opposite $y$-sides of the grain are of equal magnitude but opposite direction; thus there is no global translation force but a nonzero surface torque, leading to the behavior of pure self-rotation (CW in this case). When $\alpha_j$ increases, its asymmetric distribution along the $x$ direction with respect to the grain center results in an unbalance of surface tangential forces on the opposite $x$-side surfaces, causing a nonzero global force pointed to the perpendicular $y$ direction, together with the still nonzero but smaller net surface torque; this then leads to the behavior of self-rolling that combines $+y$ self-translation and slower CW self-rotation. A transition to the pure self-translation mode occurs when the grain center reaches $\alpha_j=0$, for which $\alpha_j$ in the two $x$-sides are of opposite sign and distribute antisymmetrically, such that net forces on the two $x$-side surfaces are equal in terms of both magnitude and direction, giving maximum translational motion but zero surface torque (no rotation). Further increasing $\alpha_j$ into the positive regime leads again to the unbalanced surface forces but CCW self-rotation, back to the phenomenon of self-rolling, until reaching the maximum location of $\alpha_j$ with its symmetric distribution across the grain and hence pure CCW self-rotation. The other half period of $\alpha_j$ variation follows the same microscopic mechanism, giving the self-traveling along the $-y$ direction and the variation from CCW to CW self-rotation, as seen in Fig.~\ref{fig:locomotion}A and Movie S16.

As also shown in Fig.~\ref{fig:locomotion}A, the time-average self-translation velocity $\langle v_c \rangle$ calculated at the steady state of simulations, with $v_c$ the velocity of centroid (geometric center) of the grain, is well fitted to a form $\langle v_c \rangle = v_{c0} \sin Q_x (x-x_0)$, while the self-rotation frequency obeys $\langle \omega \rangle = \omega_0 \cos Q_x (x-x_0)$, with $\omega>0$ for CCW rotation and $\omega<0$ for CW rotation, in phase with the $\alpha_j$ variation if compared to Eq.~\ref{eq:alphaj_1D}. This indicates a $\pi/2$ phase difference between grain self-translation and self-rotation, consistent with the above microscopic picture. It is noted that the behaviors of grain locomotion identified here are characterized by rigid-body self-motion of odd elastic crystallites, different from the usual scenarios of locomotion accompanied by body shape changes as being implemented recently in the adaptive locomotion of active metamaterials with odd elastic responses \cite{VeenstraNature25}.

Note that to obtain the above results of locomotion and mode transition, we have utilized the spatial modulation of $\alpha_1(\mathbf{r})$, in addition to that of the lowest-gradient $\alpha_0$ term. Our elasticity analysis given above (and in SI Appendix) shows that the $\alpha_0$ term itself does not generate odd elasticity, and thus the corresponding effect of grain self-rotation would not be expected. This has been verified in our simulations, where we use the same setup as that of Fig.~\ref{fig:locomotion}A other than choosing $\alpha_0^0=1$ and $\alpha_1^0=0$ in Eq.~\ref{eq:alphaj_1D} (instead of $\alpha_0^0=\alpha_1^0=1$ in Fig.~\ref{fig:locomotion}). Only grain self-translation is observed, but not self-rotation. Also, the grain becomes immobile at the smoothly varying peak and valley locations of $\alpha_0(\mathbf{r})$, as expected from the corresponding very small value of the $\bm{\nabla}(\alpha_0\psi) \times \bm{\nabla}\psi$ term in Eq.~\ref{eq:T-PFC} which then makes a negligible contribution to the grain self-motion. Thus, maintaining the $\alpha_1$ term is important to ensure effects of odd elasticity and achieve the control of transition between different locomotion modes.

These results and the mechanism unveiled enable us to steer the locomotion or transport of grains along the desired paths. This can be achieved via the pre-designed 2D spatial distribution of transverse interaction strengths. Some simulation outcomes are presented in Fig.~\ref{fig:locomotion} B and C and Movies S17 and S18, for two examples of circular and S-shaped tracks for controlled locomotion. Here we approximate a kink-type $\alpha_j$ profile separating inner (with $\alpha_j=\alpha_j^{\rm in}$) and outer (with $\alpha_j=\alpha_j^{\rm out}$) domains as
\begin{equation}
    \alpha_j(\mathbf{r}) = \frac{1}{2} \left ( \alpha_j^{\rm out} + \alpha_j^{\rm in} \right ) + \frac{1}{2} \left ( \alpha_j^{\rm out} - \alpha_j^{\rm in} \right ) \tanh \!\left ( \frac{R_s(\mathbf{r})}{\Delta} \right ),
    \label{eq:alphaj_2D}
\end{equation}
where $R_s(\mathbf{r})$ is the signed distance function from position $\mathbf{r}$ to the domain boundary (i.e., the boundary track for grain motion; see Materials and Methods) and $\Delta$ controls the boundary width. For simplicity we set $\alpha_j^{\rm in} = -\alpha_j^{\rm out} \equiv \alpha_j^0$, such that $\alpha_j=0$ at the positions of boundary. When initialized with the grain center located at the boundary track, with $+/-$ antisymmetric $\alpha_j$ distribution across the boundary, the grain will self-travel by following the exact track that has been pre-designed, as seen in Fig.~\ref{fig:locomotion} and Movies S17 and S18. The direction of grain translation is always perpendicular to the local variation direction (i.e., spatial gradient) of transverse interaction strength $\alpha_j$, based on the mechanism described above and illustrated in Fig.~\ref{fig:locomotion}A. The maneuver of locomotion modes can be expected also from Fig.~\ref{fig:locomotion}A, by placing the grain at an initial location deviated from the domain boundary, subjected to asymmetric distribution of $\alpha_j$; the grain will then locomote via self-rolling, as verified in our simulations.

\section*{Conclusions and Outlook}

We have developed a multiscale density field theory, the T-PFC model incorporating microscopic and mesoscopic length scales and diffusive timescales, for the study of complex dynamical behaviors of 2D chiral crystalline systems that are governed by both longitudinal and transverse interactions. The model incorporates nonreciprocity, 2D parity symmetry breaking, and odd elasticity through new, lowest-order field-theoretical terms originating from transverse interactions. This nonpotential continuum-field model is validated by the study of odd crystallite self-rotation, dislocation self-motion, unbinding of dislocation dipoles instead of annihilation, and motion of grain boundaries, with results well agreeing with those observed in recent experiments and particle-based simulations of various systems formed by, e.g., starfish embryos, magnetic colloids, and swimming bacteria, and consistent with the microscopic picture of transverse forces. 

Importantly, this modeling unveils a distinct type of surface cusp instability, which is the precursor of the subsequent surface dislocation nucleation and proliferation as well as the occurrence of self-fission or self-division of single-crystalline grains for strong enough transverse interaction. This behavior of grain self-fission leads to the prediction of a transition from normal to reverse Ostwald ripening, and the spontaneous formation of a dynamical polycrystalline state with its governing dynamics transitioning from the normal grain coarsening to the persistently varying process of grain self-fragmentation as effects of transverse interaction become more prominent. The critical radius of grain instability and the characteristic grain sizes in the fragmentation-dominated regimes are found to obey power-law scaling behaviors with respect to the transverse interaction strength. Also achieved is grain locomotion without shape changes as effected by net transverse interactions on grain surface, with transition between different modes of self-rotating, self-rolling, and self-translating that are controlled via local spatial variation of transverse interaction strengths. The corresponding mechanism can be used to steer the transport of individual odd grain along designed trajectories that are either straight or curved, an intriguing feature for odd elastic systems.

The T-PFC model developed here constitutes a bridging between particle-based microscopic picture and continuum odd elasticity theory, enabling the simulations across a broad range of scales and system sizes that are of experimental relevance (ranging from a few to tens of thousands of particles), as demonstrated in the above calculations. Much larger systems can be simulated when needed, without losing particle resolution, through this T-PFC modeling and also the corresponding mesoscopic amplitude equations (Eqs.~\ref{eq:psi0} and \ref{eq:Aj}) as has been well demonstrated in previous PFC studies of passive systems. The modeling can be applied to examine and predict much more complex phenomena in chiral, odd crystalline systems that are beyond those presented here, such as mechanical and dynamical properties subjected to various external conditions, plastic deformation and dynamics of topological defects under variations of transverse interaction strengths, effects of confinements with different geometries and topologies, among many others. An interesting application for models of this type would be magnetic skyrmions, another example for a system with transverse interactions \cite{KalzVSMS2024,ReichhardtRM2022,EverschorMRK2018} and in particular odd elasticity \cite{BravermanPRL21}. In addition, the model is readily extended to the study of self-propelled spinners \cite{ChenNatPhys25} and multicomponent mixtures with transverse interactions within the same and between different species, which are expected to generate a rich variety of complex dynamical behaviors and patterns as a result of coupling and competition between different degrees of freedom and multiple scales that can be well described by this PFC-type continuum approach.

\section*{Materials and Methods}

\subsection*{Numerical Simulations}
The T-PFC model equation \ref{eq:T-PFC} or \ref{eq:TPFC} is solved numerically via the pseudospectral method with the use of periodic boundary conditions. A corresponding code is available in a GitHub repository \cite{Code}. The model parameters are chosen as $\epsilon=0.1$, $g=0.5$, $q_0=1$, and $\alpha_2=0$, with different values of $\alpha_1$ and average density $\bar{\psi}_0$. When studying the cases with solid-liquid coexistence, $\bar{\psi}_0$ values in solid and liquid regions are chosen and adjusted according to the phase diagram calculated from original PFC \cite{Provatas10} given that in one-mode approximation effects of transverse interaction can be neglected in the perfect bulk state. Sizes of simulated systems range from $512 \times 512$ to $4096 \times 4096$ grid points, with grid spacing $\Delta x = \Delta y = \pi/4$ and simulation time step $\Delta t = 0.5$ (with similar results obtained if using smaller $\Delta t$). Thus each system simulated for the study of multigrain dynamics in the polycrystalline state shown in Fig.~\ref{fig:multigrains} (of size $2048 \times 2048$ up to $t=10^5$) contains roughly $10^5$ particles or density peaks. More details of simulation setups are given in SI Appendix.

For the study of grain locomotion, the simulated system size is set as $1024 \times 512$ grid points in Fig.~\ref{fig:locomotion}A (with $\bar{\psi}_0|_{\rm solid}=-0.076$ and $\bar{\psi}_0|_{\rm liquid}=-0.116$) and $512 \times 512$ in Fig.~\ref{fig:locomotion} B and C (with $\bar{\psi}_0|_{\rm solid}=-0.078$ and $\bar{\psi}_0|_{\rm liquid}=-0.115$). The parameters used for the setup of $\alpha_j$ spatial variations are $\alpha_0^0=\alpha_1^0=1$, $Q_x = 2\pi/L_x$ with $L_x=1024\Delta x$, $x_0=9L_x/16$, and $\Delta = 8\Delta x$. For the circular trajectory shown in Fig.~\ref{fig:locomotion}B, the signed distance function $R_s$ in Eq.~\ref{eq:alphaj_2D} of the corresponding $\alpha_j(\mathbf{r})$ profile is given by
\begin{equation}
    R_s(\mathbf{r}) = |\mathbf{r}-\mathbf{r}_c| - r_0,
\end{equation}
with center position $\mathbf{r}_c$ and radius $r_0$ of the circle. For the S-shaped channel with two boundary tracks shown in Fig.~\ref{fig:locomotion}C, the corresponding $R_s$ in Eq.~\ref{eq:alphaj_2D} is approximated by
\begin{equation}
    R_s(\mathbf{r}) \approx |x-x_c| - \left [ r_0 \mp S_0 \sin (Q_s(y-y_c)) \right ],
\end{equation}
where $(x_c, y_c)$ is the location of the channel center, $r_0$ is the half width of the channel, $S_0$ and $Q_s$ give the amplitude and wave number of the periodic S-shaped modulation, and ``$-$'' (``$+$'') is used for the region of $x > x_c$ ($x < x_c$). A variety of other types of $\alpha_j$ spatial profiles, with the corresponding boundary tracks set for different geometries or topologies, can be set up in a similar way through $R_s(\mathbf{r})$ (see, e.g., Ref.~\cite{HuangCP22}).

\section*{Acknowledgments}
We thank Tzer Han Tan for providing the experimental data of Ref.~\cite{TanNature22} for cluster rotation frequencies of starfish embryos. Z.-F.H. acknowledges support from the National Science Foundation under Grant No. DMR-2006446. M.t.V.,  R.W., and H.L. are funded by the Deutsche Forschungsgemeinschaft (DFG, German Research Foundation) -- Project-IDs  465145163 -- SFB 1552 (M.t.V.), 535275785 (R.W.), and LO 418/29-1 (H.L.).

\bibliography{T-PFC_refs}

\end{document}


\title{Supporting Information for \texorpdfstring{\\}{}
  Anomalous grain dynamics and grain locomotion of odd crystals}

\author{Zhi-Feng Huang}
\affiliation{Department of Physics and Astronomy, Wayne State University, Detroit, Michigan 48201, USA}
\author{Michael te Vrugt}
\affiliation{Institut f\"{u}r Physik, Johannes Gutenberg-Universit\"{a}t Mainz, 55128 Mainz, Germany}
\author{Raphael Wittkowski}
\affiliation{Department of Physics, RWTH Aachen University, 52074 Aachen, Germany}
\affiliation{DWI -- Leibniz Institute for Interactive Materials, 52074 Aachen, Germany}
\author{Hartmut L\"{o}wen}
\affiliation{Institut f\"{u}r Theoretische Physik II: Weiche Materie, Heinrich-Heine-Universit\"{a}t D\"{u}sseldorf, 40225 D\"{u}sseldorf, Germany}

\maketitle

\section*{Model derivation}

\subsection*{Transverse interaction}
We start from the microscopic equations of motion for $N$ Brownian particles with transverse interactions in two spatial dimensions (2D). The $i$-th particle, located at position $\bm{r}_i$, undergoes overdamped motion and experiences, from the $j$-th particle, a conservative longitudinal force $\mathbf{F}_{ij}^\mathrm{c}$ and a nonconservative transverse force $\mathbf{F}_{ij}^\perp$. It is thereby governed by the Langevin equations
\begin{equation}
\dot{\bm{r}}_i(t)= \beta D_\mathrm{T} \sum_{j\neq i}(\mathbf{F}_{ij}^\mathrm{c}+\mathbf{F}_{ij}^\perp) + \sqrt{2D_\mathrm{T}}\bm{\xi}_i,
\label{langevinequations}
\end{equation}
where $\beta = 1/(k_\mathrm{B}T)$ with Boltzmann constant $k_\mathrm{B}$ and temperature $T$ is the thermodynamic beta, $D_\mathrm{T}$ is the translational diffusion coefficient, and $\bm{\xi}_i$ is a white noise with zero mean and unit variance. In two spatial dimensions (2D), the interparticle transverse force $\mathbf{F}_{ij}^\perp$ is of the general form
\begin{equation}
\mathbf{F}_{ij}^\perp=f_{ij}^\perp \, \bm{\hat{z}}\times \frac{\bm{r}_{ij}}{r_{ij}},
\label{eq:f_ij}
\end{equation}
with the interparticle relative position vector $\bm{r}_{ij} = \bm{r}_i - \bm{r}_j$, $r_{ij}=\norm{\bm{r}_{ij}}$, and the unit vector $\bm{\hat{z}}$ in the $z$ direction. We write the position vector as $\mathbf{r}_{ij} = r_{ij} \bm{\hat{u}}(\phi_{\mathrm{R}})$, where $\bm{\hat{u}}$ in 2D is parametrized via an angle $\phi_{\mathrm{R}}$ in the form $\bm{\hat{u}}=(\cos(\phi_{\mathrm{R}}),\sin(\phi_{\mathrm{R}}))^\mathrm{T}$. The transverse interaction force in Eq.\ \ref{eq:f_ij} then reads
\begin{equation}
\bm{F}_{ij}^\perp=f^\perp\bigg(\frac{\mathbf{r}_j+\mathbf{r}_j}{2},\mathbf{r}_i-\mathbf{r}_j,t\bigg)
\begin{pmatrix}
-\sin(\phi_{\mathrm{R}})\\
\cos(\phi_{\mathrm{R}})
\end{pmatrix}.
\label{eq:force}
\end{equation}
Note that we allow here the force function $f^\perp$ to explicitly depend not only on the separation vector $\mathbf{r}_i-\mathbf{r}_j$, but also on the center of mass position\footnote{We assume that all particles have equal mass.} $(\mathbf{r}_i+\mathbf{r}_j)/2$ and on time $t$; i.e., it is possible that the transverse interaction has different magnitudes at different locations in space. This is exploited for the control of grain locomotion in the main text.

The Smoluchowski equation corresponding to the Langevin equations \ref{langevinequations} is given by
\begin{equation}
\pdif{}{t}\Psi = \sum_{i=1}^{N}\Big[D_\mathrm{T}\bm{\nabla}_i^2\Psi - \beta D_\mathrm{T}\bm{\nabla}_i\cdot(\bm{F}_{ij}^\mathrm{c}\Psi) - \beta D_\mathrm{T}\bm{\nabla}_i\cdot(\bm{F}_{ij}^\perp\Psi)\Big],
\label{smoluchowski}
\end{equation}
where $\Psi(\bm{r}_1,\dotsb,\bm{r}_N,t)$ is the $N$-body phase-space probability distribution function and $\bm{\nabla}_i$ is the gradient operator with respect to the position vector $\bm{r}_i$. We define the one-body and two-body densities as
\begin{align}
\rho(\bm{r},t)&=N\INT{}{}{^2r_2}\dotsc\INT{}{}{^2r_N}\Psi(\bm{r},\bm{r}_2,\dotsb,\bm{r}_N,t),\\
\rho^{(2)}(\bm{r},\bm{r}',t)&=N(N-1)\INT{}{}{^2r_3}\dotsc\INT{}{}{^2r_N}\Psi(\bm{r},\bm{r}',\bm{r}_3,\dotsb,\bm{r}_N,t),
\end{align}
and write the conservative longitudinal force as
\begin{equation}
\bm{F}_{ij}^\mathrm{c} = - \bm{\nabla}_i U_2(\bm{r}_i-\bm{r}_j),
\end{equation}
where $U_2$ is the interaction potential between two particles. Then, integrating Eq.\ \ref{smoluchowski} over the coordinates of all particles except for one gives 
\begin{equation}
\pdif{}{t}\rho(\bm{r},t) = D_\mathrm{T}\bm{\nabla}^2\rho(\bm{r},t) + \beta D_\mathrm{T}\bm{\nabla}\cdot\bigg(\INT{}{}{^2r'}\rho^{(2)}(\bm{r},\bm{r}',t)\bm{\nabla}U_2(\bm{r}-\bm{r}')\bigg) - \beta D_\mathrm{T}\bm{\nabla}\cdot\bigg(\INT{}{}{^2r'}\rho^{(2)}(\bm{r},\bm{r}',t)\bm{F}^\perp(\bm{r},\bm{r}',t)\bigg).
\label{smoluchowskireduced}
\end{equation}
The first two terms in Eq.\ \ref{smoluchowskireduced} are standard and can be treated via the usual approximations of dynamical density functional theory. This procedure was outlined in Refs. \cite{archer2004dynamical,teVrugtLW2020} and is not repeated here. It yields
\begin{equation}
\pdif{}{t}\rho(\bm{r},t) = \beta D_\mathrm{T} \bm{\nabla}\cdot\bigg(\rho(\bm{r},t)\bm{\nabla}\Fdif{F}{\rho(\bm{r},t)}\bigg) - \beta D_\mathrm{T}\bm{\nabla}\cdot\bigg(\INT{}{}{^2r'}\rho^{(2)}(\bm{r},\bm{r}',t)\bm{F}^\perp(\bm{r},\bm{r}')\bigg)
\label{smoluchowskireduced2}
\end{equation}
with the free energy functional $F$ which will be specified later. For the remaining third term of Eq.\ \ref{smoluchowskireduced}, we make a mean-field approximation to get
\begin{equation}
\frac{\partial}{\partial t} \rho(\mathbf{r},t) 
    = \beta D_\mathrm{T} \bm{\nabla}\cdot\bigg(\rho(\bm{r},t)\bm{\nabla}\Fdif{F}{\rho(\bm{r},t)}\bigg)
       -  \beta D_\mathrm{T}  \bm{\nabla}\cdot
            \bigg(
                \rho(\mathbf{r},t) \INT{}{}{^2 r'}
                \Big[
                \mathbf{F}^\perp(\mathbf{r},\mathbf{r}',t) \rho(\mathbf{r}',t)
                \Big]
            \bigg).
   \label{eq:DDFT_raw}
\end{equation}
For the first term in Eq.\ \ref{eq:DDFT_raw}, we make a lowest-order approximation via scaling analysis \cite{HuangPRE10} that is equivalent to the constant mobility approximation \cite{ArcherRRS2019}
\begin{equation}
\bm{\nabla}\cdot\bigg(\rho(\bm{r},t)\bm{\nabla}\Fdif{F}{\rho(\bm{r},t)}\bigg)
\approx \rho_0 \bm{\nabla}^2\Fdif{F}{\rho(\bm{r},t)},
\label{eq:cma}
\end{equation}
with a reference-state density $\rho_0$. For the second term in Eq.\ \ref{eq:DDFT_raw}, we remove the nonlocality via a gradient expansion \cite{YangFG1976}. Specifically, we first make the substitution
\begin{equation}
\mathbf{r}'\to\mathbf{r}+\mathbf{r}'.
\label{eq:substitution}
\end{equation}
Using Eq.\ \ref{eq:force}, the second term in Eq.\ \ref{eq:DDFT_raw} then reads
\begin{equation}
\beta D_\mathrm{T} \partial_i
            \bigg(
                \rho(\mathbf{r},t) \INT{}{}{^2 r'}
                \Big[
                f^\perp\bigg(\frac{\mathbf{r}'+2\mathbf{r}}{2},\mathbf{r}',t\bigg)\epsilon_{ij}u_j(\varphi_\mathrm{R}) \rho(\mathbf{r}+\mathbf{r}',t)
                \Big]
            \bigg),
\label{eq:secondterm}
\end{equation}
with summation over spatial indices\footnote{From now on, we use $i$ and $j$ for spatial indices rather than for particle labels.} appearing twice (i.e., the Einstein summation convention) and the 2D Levi-Civita symbol
\begin{equation}
\mathbf{\epsilon}=
\begin{pmatrix}
 0 &  1\\
 -1 & 0
\end{pmatrix}.
\end{equation}
For $\rho(\mathbf{r}+\mathbf{r}',t)$, we use the gradient expansion \cite{teVrugtBW2022,ArcherRRS2019}
\begin{equation}
\rho(\mathbf{r}+\mathbf{r}',t)= \sum_{l=0}^{\infty}\frac{R^l}{l!}(u_{k}(\phi_{R})\partial_{k})^l\rho(\mathbf{r},t).\label{gradientexpansion}
\end{equation}
The physical reasoning behind this gradient expansion -- and in particular behind a truncation at a finite $l$ -- is the assumption that the interaction force is sufficiently short-ranged, i.e., $f^\perp$ vanishes for large enough $R$ (where $R$ is the modulus of $\mathbf{r}'$). We now assume that
\begin{enumerate}
    \item the dependence of $f^\perp$ on $\mathbf{r}'$ can be replaced by a dependence on $R$; in other words, $f^\perp$ does not depend on the spatial orientation of the vector $\mathbf{r}'$;
    \item the spatial modulation of $f^\perp$, which is encoded in the first argument of $f^\perp$, occurs only on length scales substantially larger than the range of the interaction force, corresponding to situations in most experimental systems.
\end{enumerate}
The second assumption implies that we can, to a very good approximation, replace $(\mathbf{r}'+2\mathbf{r})/\mathbf{r}$ by $\mathbf{r}$ in the first argument of $f^\perp$. Using Eq.\ \ref{gradientexpansion}, Eq.\ \ref{eq:secondterm} then simplifies to 
\begin{equation}
\beta D_\mathrm{T}\sum_{l=0}^{\infty} \partial_i\Big[\rho(\mathbf{r},t) \INT{0}{\infty}{R}\INT{0}{2\pi}{\phi_\mathrm{R}}
                \frac{R^{l+1}}{l!}f^\perp(\mathbf{r},R,t)\epsilon_{ij}u_j(\varphi_\mathrm{R}) (u_k(\phi_{R})\partial_k)^l\rho(\mathbf{r},t)\Big].
\label{eq:secondterm2}
\end{equation}
Combining Eqs.\ \ref{eq:DDFT_raw}, \ref{eq:cma}, and \ref{eq:secondterm2} gives
\begin{align}
\frac{\partial}{\partial t} \rho(\mathbf{r},t) 
    =\,&  \beta D_\mathrm{T} \rho_0 \bm{\nabla}^2 \Fdif{F}{\rho(\mathbf{r},t)}
        + \beta D_\mathrm{T} \sum_{l=0}^{\infty} \partial_i
            \bigg[
                \rho(\mathbf{r},t) \INT{}{}{R}\INT{0}{2\pi}{\phi_{\mathrm{R}}}\frac{R^{l+1}}{l!}f^\perp(\mathbf{r},R,t)\epsilon_{ik}u_k(\phi_{\mathrm{R}})(u_j(\phi_{\mathrm{R}})\partial_j)^l \rho(\mathbf{r},t)
            \bigg].
   \label{eq:DDFT_raw2}
\end{align}
In the following we will not write explicitly the dependence on $\bm{r}$ and $t$ unless needed. Integrating out Eq.\ \ref{eq:DDFT_raw2} and truncating at $l=5$ gives
\begin{equation}
\frac{\partial \rho}{\partial t} 
    =D_\mathrm{T} \left \{ \beta \rho_0 \bm{\nabla}^2 \Fdif{F}{\rho}
      +\partial_i \left [ \rho\epsilon_{ij} \left ( \mathcal{A}_0 \partial_j\rho + \mathcal{A}_1 \partial_k^2\partial_j \rho+\mathcal{A}_2 (\partial_k^2)^2\partial_j \rho \right ) \right ] \right \},
      \label{eq:rho_transverse}
\end{equation}
with the coefficients
\begin{align}
\mathcal{A}_0(\mathbf{r},t)&=\pi\beta\INT{0}{\infty}{R}R^2f^\perp(\mathbf{r},R,t), \label{eq:A0}\\
\mathcal{A}_1(\mathbf{r},t)&=\frac{\pi}{8}\beta\INT{0}{\infty}{R}R^4f^\perp(\mathbf{r},R,t), \label{eq:A1}\\
\mathcal{A}_2(\mathbf{r},t)&=\frac{\pi}{192}\beta\INT{0}{\infty}{R}R^6f^\perp(\mathbf{r},R,t) \label{eq:A2}.
\end{align}
Here only terms of odd order of $l$ in the gradient expansion contribute to Eq.~\ref{eq:rho_transverse}, since
\begin{align}
&\INT{0}{2\pi}{\phi_\mathrm{R}}u_j(\phi_\mathrm{R})=0,\label{eq:onlyodd1}\\
&\INT{0}{2\pi}{\phi_\mathrm{R}}u_j(\phi_\mathrm{R})(u_k(\phi_\mathrm{R})\partial_k)^2=0,\label{eq:onlyodd2}\\
&\INT{0}{2\pi}{\phi_\mathrm{R}}u_j(\phi_\mathrm{R})(u_k(\phi_\mathrm{R})\partial_k)^4=0.
\end{align}
This leads to the appearance of even-order gradients in Eq.~\ref{eq:rho_transverse}.
Note that the coefficients $\mathcal{A}_0, \mathcal{A}_1, \mathcal{A}_2$ are time- and position-dependent, i.e., $\mathcal{A}_j=\mathcal{A}_j(\mathbf{r},t)$, in the case of spatially and/or temporally varying transverse force $f^\perp$. Since $\epsilon_{1j} \partial_j = \partial_y$ and $\epsilon_{2j} \partial_j = -\partial_x$, Eq.\ \ref{eq:rho_transverse} can be rewritten as
\begin{align}
    \frac{\partial \rho}{\partial t} &= D_\mathrm{T} \left \{ \beta \rho_0 \bm{\nabla}^2 \Fdif{F}{\rho} + \left [ \partial_x (\mathcal{A}_0\rho) \right ] \partial_y \rho - \left [ \partial_y (\mathcal{A}_0\rho) \right ] \partial_x \rho \right. \nonumber\\
    & \qquad\; \left. + \left [ \partial_x (\mathcal{A}_1\rho) \right ] \partial_y \nabla^2 \rho - \left [ \partial_y (\mathcal{A}_1\rho) \right ] \partial_x \bm{\nabla}^2 \rho + \left [ \partial_x (\mathcal{A}_2\rho) \right ] \partial_y \bm{\nabla}^4 \rho - \left [ \partial_y (\mathcal{A}_2\rho) \right ] \partial_x \bm{\nabla}^4 \rho \right \},
\end{align}
such that in 2D
\begin{equation}
    \frac{\partial \rho}{\partial t} = D_\mathrm{T} \left \{ \beta \rho_0 \bm{\nabla}^2 \Fdif{F}{\rho} + \left [ \bm{\nabla} (\mathcal{A}_0\rho) \times \bm{\nabla} \rho + \bm{\nabla} (\mathcal{A}_1\rho) \times \bm{\nabla} \bm{\nabla}^2 \rho + \bm{\nabla} (\mathcal{A}_2\rho) \times \bm{\nabla} \bm{\nabla}^4 \rho \right ]_z \right \}.
    \label{eq:transverse}
\end{equation}
If the $\mathcal{A}_{j=0,1,2}$ coefficients are spatially constant, Eq.\ \ref{eq:transverse} simplifies to
\begin{equation}
    \frac{\partial \rho}{\partial t} = D_\mathrm{T} \left \{ \beta \rho_0 \bm{\nabla}^2 \Fdif{F}{\rho} + \left [ \left ( \bm{\nabla} \rho \right ) \times \bm{\nabla} \left ( \mathcal{A}_1 \bm{\nabla}^2 \rho + \mathcal{A}_2 \bm{\nabla}^4 \rho \right ) \right ]_z \right \}.
\end{equation}
The expressions of $\mathcal{A}_j$ depend on the specific form of the transverse interaction. For example, if using the form of hydrodynamic near-field interaction force obtained from the lubrication theory, as applied to the study of chiral living crystals of starfish embryos, we have \cite{TanNature22}
\begin{equation}
f^\perp(r)=
\begin{cases}
&f_0^\perp\ln\big(\frac{r_\mathrm{c}}{r - 2r_0}\big), \text{ for } 2r_0 < r < 2r_0 + r_\mathrm{c}\\
&0, \text{ otherwise}
\end{cases}
\label{eq:starfish}
\end{equation}
where $r_0$ is the radius of the spinning particle and $r_c$ sets the cutoff distance of the transverse force. Substituting Eq.\ \ref{eq:starfish} into Eqs.~\ref{eq:A0}--\ref{eq:A2} leads to the explicit expressions
\begin{align}
\mathcal{A}_0&=  \frac{\pi}{9}f_0^\perp\beta r_\mathrm{c} (36 r_0^2+9r_0 r_\mathrm{c} + r_\mathrm{c}^2), \label{eq:A0_hydro}\\ 
\mathcal{A}_1&= \frac{\pi}{8}f_0^\perp\beta \bigg(16 r_0^4r_\mathrm{c}+8r_0^3r_\mathrm{c}^2 + \frac{8r_0^2r_\mathrm{c}^3}{3} + \frac{r_0r_\mathrm{c}^4}{2}+\frac{r_\mathrm{c}^5}{25}\bigg), \label{eq:A1_hydro}\\ 
\mathcal{A}_2&= \frac{\pi}{192}f_0^\perp\beta\bigg(64 r_0^6r_\mathrm{c} + 48 r_0^5r_\mathrm{c}^2+\frac{80r_0^4r_\mathrm{c}^3}{3}+10r_0^3r_\mathrm{c}^4 + \frac{12r_0^2r_\mathrm{c}^5}{5}+\frac{r_0r_\mathrm{c}^6}{3}+\frac{r_\mathrm{c}^7}{49}\bigg). \label{eq:A2_hydro}
\end{align}
This is an example for the case where the $\mathcal{A}_j$ coefficients are spatially independent constants.

We now briefly examine what would happen if the spatial modulation of $f^\perp$ is not weak on scales comparable to its range, i.e., varies on small length scales (although this scenario is difficult to realize in experimental systems). In this case, we would need to do a gradient expansion also for $f^\perp$, which reads
\begin{equation}
f^\perp\bigg(\frac{\mathbf{r}'+2\mathbf{r}}{2},R,t\bigg)= \sum_{m=0}^{\infty}\frac{R^m}{2^m m!}(u_{k}(\phi_{R})\partial_{k})^m f^\perp(\mathbf{r},R,t).\label{gradient_expansion}
\end{equation}
Equation \ref{eq:DDFT_raw2} then becomes
\begin{align}
\frac{\partial}{\partial t} \rho(\mathbf{r},t) 
    =&~ \beta D_\mathrm{T}  \rho_0 \bm{\nabla}^2 \Fdif{F}{\rho(\mathbf{r},t)} \nonumber\\
     & + \beta D_\mathrm{T}\sum_{l,m=0}^{\infty} \partial_i
            \bigg[
                \rho(\mathbf{r},t) \INT{}{}{R}\INT{0}{2\pi}{\phi_{\mathrm{R}}}\frac{R^{l+m+1}}{2^m m!l!}((u_n(\phi_{\mathrm{R}})\partial_n)^m f^\perp(\mathbf{r},R,t))\epsilon_{ik}u_k(\phi_{\mathrm{R}})(u_j(\phi_{\mathrm{R}})\partial_j)^l \rho(\mathbf{r},t)
            \bigg].
   \label{eq:DDFT_raw2alt}
\end{align}
Dropping all terms with $l+m>3$ (in other words, dropping terms of higher than fourth order in gradients) and evaluating the integrals give
\begin{equation}
\begin{split}
\frac{\partial \rho}{\partial t} 
=&~D_\mathrm{T} \bigg\{ \beta \rho_0 \bm{\nabla}^2 \Fdif{F}{\rho}+\partial_i \bigg[ \rho\epsilon_{ij} \bigg({\mathcal{A}_0} \partial_j\rho + \frac{1}{2}\rho \partial_j {\mathcal{A}_0} \\
  &\qquad + {\mathcal{A}_1}\partial_k^2\partial_j\rho + \frac{1}{2}\rho\partial_k^2\partial_j {\mathcal{A}_1}+ \frac{1}{2}(\partial_j\rho) \partial_k^2 {\mathcal{A}_1} + \frac{1}{2}(\partial_j {\mathcal{A}_1})\partial_k^2\rho +(\partial_k\rho)\partial_j \partial_k {\mathcal{A}_1} + (\partial_k {\mathcal{A}_1})\partial_j \partial_k\rho\bigg) \bigg] \bigg \}.    
\end{split}
\label{eq:rho_transversealt}
\end{equation}
Even though here we have dropped terms of sixth order in gradients, Eq.\ \ref{eq:rho_transversealt} is considerably more complicated than Eq.\ \ref{eq:rho_transverse} with higher-order derivatives of $\mathcal{A}_j$. As a consequence of Eqs.\ \ref{eq:onlyodd1} and \ref{eq:onlyodd2}, there are no contributions with $l+m=0$ and $l+m=2$. Note that as in Eq.~\ref{eq:rho_transverse}, all the $\mathcal{A}_j$ terms originated from transverse interactions are of a purely nonpotential form.

In this study we focus on a minimal field-theoretical model that incorporates and produces all the key properties of odd crystals, and thus use the realistic and simpler version of Eq.~\ref{eq:rho_transverse} or \ref{eq:transverse}, instead of the complicated case of Eq.~\ref{eq:rho_transversealt}.

\subsection*{T-PFC model}

For the free energy $F$, we use the simplest approximation that allows to model the crystalline phases, namely the phase field crystal (PFC) free energy functional $F_{\rm PFC}$ \cite{Provatas10,EmmerichAdvPhys12}. This incorporates the effects of longitudinal interactions. The original form with one microscopic lattice length scale (of characteristic wave number $Q_0$) is given by \cite{Provatas10}
\begin{equation}
    F_{\rm PFC} = \int d\mathbf{r} \left \{ \frac{1}{2} \phi \left [ \Delta \mathcal{B} + \lambda \left ( \bm{\nabla}^2 + Q_0^2 \right )^2 \right ] \phi - \frac{1}{3} \tau \phi^3 + \frac{1}{4}u \phi^4 \right \},
    \label{eq:Fpfc}
\end{equation}
and the corresponding conserved dynamics $D_{\rm T} \bm{\nabla}^2 \delta F_{\rm PFC} / \delta \phi$ can be incorporated into Eq.~\ref{eq:transverse}. Similarly we could use other forms of $F_{\rm PFC}$, such as those for multi-model PFC \cite{MkhontaPRL13} or bond-angle dependent PFC \cite{WangPRB18} or others, to model a variety of crystalline lattice symmetries. Here, $F_{\rm PFC}$ corresponds to $\beta F$ \cite{HuangPRE10}, and the density variation field $\phi = (\rho - \rho_0) / \rho_0$, with $\rho_0$ the reference-state density. The polynomial terms with coefficients $\Delta \mathcal{B}$, $\tau$, and $u$ in Eq.~\ref{eq:Fpfc} are analogous to those of standard Landau theory giving two asymmetric potential wells (as a result of nonzero cubic term) that lead to first-order phase transition. Importantly, the quadratic term with a gradient form $(\nabla^2 + Q_0^2)^2$ yields a liquid-state structure factor peaked at the wave number $Q_0$ \cite{ElderPRE04}, and the minimization of the corresponding free energy in the solid state (with $\Delta \mathcal{B} > 0$) results in a spatially periodic solution of the density field $\phi$ characterized by the same $Q_0$ (representing a single lattice length scale of 2D triangular structure) \cite{Provatas10}. It has also been shown that all these PFC terms can be derived from classical density functional theory (cDFT) of freezing \cite{ElderPRB07} or dynamical DFT (DDFT) \cite{HuangPRE10}, such that those PFC coefficients can be expressed in terms of the expansion components of two- and three-point direct correlation functions in Fourier space, i.e. \cite{HuangPRE10},
\begin{equation}
    \Delta \mathcal{B} = 1 + \rho_0 \left [ \hat{C}_0 - \hat{C}_2^2/(4\hat{C}_4) \right ], \quad \lambda = \rho_0 \hat{C}_4, \quad Q_0 = \hat{C}_2/(2\hat{C}_4), \quad \tau = \frac{1}{2} \rho_0 \left [ \hat{C}_2^2/(4\hat{C}_4) - \hat{C}_0 \right ] - \frac{1}{2} \rho_0^2 \hat{C}_0^{(3)}, \quad u = \frac{1}{3} \rho_0^2 \hat{C}_0^{(3)},
    \label{eq:coefficients_DDFT}
\end{equation}
given the expansion $\hat{C}^{(2)}(q) = -\hat{C}_0 + \hat{C}_2 q^2 - \hat{C}_4 q^4 + \cdots$ and $\hat{C}^{(3)}(\mathbf{q},\mathbf{q}') \approx \hat{C}^{(3)}(\mathbf{q}=\mathbf{q}'=0) = -\hat{C}_0^{(3)}$, where $\hat{C}^{(2)}(q)$ and $\hat{C}^{(3)}(\mathbf{q},\mathbf{q}')$ are Fourier transforms of the liquid-state two- and three-point direct correlation functions $C^{(2)}$ and $C^{(3)}$, respectively.

Note that this form of PFC model presented in Eq.~\ref{eq:Fpfc} looks similar to that of the classical Swift-Hohenberg model \cite{Provatas10}, but with two differences. The first one is that a nonzero cubic term of $\phi^3$ is incorporated explicitly in this PFC free energy functional but not in the original Swift-Hohenberg model which gives stripe patterns. The second difference is crucial, that is, the corresponding model dynamics is conserved in PFC but nonconserved in the Swift-Hohenberg model. This conserved PFC dynamics leads to time-independent constant average density $\bar{\phi}$ and makes the first difference not essential. If writing the density field as $\phi = \bar{\phi} + \tilde{\phi}$ with variation $\tilde{\phi}$, an effective cubic term of $\tilde{\phi}^3$ will appear from the quartic $\phi^4$ term, then breaking the even symmetry of the free energy, as needed to obtain the first-order liquid-solid phase transition including phase coexistence and metastability. Thus, compared to the free energy form with only a quartic $\phi^4$ term (in addition to the quadratic terms) as appearing in the original formulation of the PFC model \cite{ElderPRL02,ElderPRE04}, adding an additional $\phi^3$ term would just shift the phase boundaries with respect to $\bar{\phi}$ in the liquid-solid phase diagram, making different versions of PFC models with and without the cubic term in principle equivalent. (This is different from the Swift-Hohenberg model with nonconserved dynamics, where any initial nonzero average density $\bar{\phi}$ varies with time and decays to zero, such that $\phi^4=\tilde{\phi}^4$, i.e., the quartic term itself cannot lead to an effective cubic one and thus no metastability occurs in the Swift-Hohenberg model with only a $\phi^4$ term which generates 2D stripes but not triangular crystalline structure.) Considering also that the $\phi^3$ term appears during the derivation of the PFC model from cDFT or DDFT (see Eq.~\ref{eq:coefficients_DDFT}), here we include this cubic term in the PFC free energy functional, as in recent PFC studies \cite{Provatas10}, to explicitly indicate that PFC models incorporate the properties of first-order phase transition and metastability.

Incorporating the effects of transverse interactions, from Eqs.~\ref{eq:transverse} and \ref{eq:Fpfc} the T-PFC model dynamics is governed by
\begin{align}
    \frac{\partial \phi}{\partial t} =&~ D_\mathrm{T} \left \{ \left [ \bm{\nabla} \left (\tilde{\mathcal{A}}_0 (1+\phi) \right ) \times \bm{\nabla} \phi + \bm{\nabla} \left (\tilde{\mathcal{A}}_1 (1+\phi) \right ) \times \bm{\nabla} \bm{\nabla}^2 \phi + \bm{\nabla} \left (\tilde{\mathcal{A}}_2 (1+\phi) \right ) \times \bm{\nabla} \bm{\nabla}^4 \phi \right ]_z \right. \nonumber\\
    &~ \left. + \bm{\nabla}^2 \left [ \Delta \mathcal{B}~\phi + \lambda \left ( \bm{\nabla}^2 + Q_0^2 \right )^2 \phi - \tau \phi^2 + u \phi^3 \right ] \right \},
\end{align}
where $\tilde{\mathcal{A}}_{j} = \rho_0 \mathcal{A}_{j}$ ($j=0,1,2$). Setting $\tilde{\psi} = 1+\phi = \rho/\rho_0$ leads to
\begin{align}
    \frac{\partial \tilde{\psi}}{\partial t} =&~ D_\mathrm{T} \left \{ \left [ \bm{\nabla} \left (\tilde{\mathcal{A}}_0 \tilde{\psi} \right ) \times \bm{\nabla} \tilde{\psi} + \bm{\nabla} \left (\tilde{\mathcal{A}}_1 \tilde{\psi} \right ) \times \bm{\nabla} \bm{\nabla}^2 \tilde{\psi} + \bm{\nabla} \left (\tilde{\mathcal{A}}_2 \tilde{\psi} \right ) \times \bm{\nabla} \bm{\nabla}^4 \tilde{\psi} \right ]_z \right. \nonumber\\
    &~ \left. + \bm{\nabla}^2 \left [ \Delta \mathcal{B}' \tilde{\psi} + \lambda \left ( \bm{\nabla}^2 + Q_0^2 \right )^2 \tilde{\psi} - \tau' \tilde{\psi}^2 + u \tilde{\psi}^3 \right ] \right \},
\end{align}
where $\Delta \mathcal{B}' = \Delta \mathcal{B} + 2\tau + 3u$ and $\tau' = \tau+3u$.

Rescaling with respect to a length scale $Q_0^{-1}$, a time scale $(D_\mathrm{T} \lambda Q_0^6)^{-1}$, and $\tilde{\psi} \rightarrow \sqrt{\lambda Q_0^4 /u}~\psi$, we reach a dimensionless PFC equation incorporating transverse interactions (T-PFC)
\begin{equation}
    \frac{\partial \psi}{\partial t} = \left [ \bm{\nabla} \left (\alpha_0 \psi \right ) \times \bm{\nabla} \psi + \bm{\nabla} \left (\alpha_1 \psi \right ) \times \bm{\nabla} \bm{\nabla}^2 \psi + \bm{\nabla} \left (\alpha_2 \psi \right ) \times \bm{\nabla} \bm{\nabla}^4 \psi \right ]_z + \bm{\nabla}^2 \left [ -\epsilon \psi + \left ( \bm{\nabla}^2 + q_0^2 \right )^2 \psi - g\psi^2 + \psi^3 \right ],
    \label{eq:T-PFC}
\end{equation}
where
\begin{align}
    & \alpha_0 = \frac{\tilde{\mathcal{A}}_0}{\sqrt{u\lambda Q_0^4}} = \frac{\rho_0}{\sqrt{u\lambda} Q_0^2} \mathcal{A}_0, \qquad
    \alpha_1 = \frac{Q_0^2 \tilde{\mathcal{A}}_1}{\sqrt{u\lambda Q_0^4}} = \frac{\rho_0}{\sqrt{u\lambda}} \mathcal{A}_1, \qquad
    \alpha_2 = \frac{Q_0^4 \tilde{\mathcal{A}}_2}{\sqrt{u\lambda Q_0^4}} = \frac{\rho_0 Q_0^2}{\sqrt{u\lambda}} \mathcal{A}_2, \label{eq:rescaling1}\\
    & \epsilon = -\frac{\Delta \mathcal{B}'}{\lambda Q_0^4} = -\frac{\Delta \mathcal{B} + 2\tau + 3u}{\lambda Q_0^4}, \qquad
    q_0 = 1, \qquad
    g = \frac{\tau'}{\sqrt{u\lambda} Q_0^2} = \frac{\tau + 3u}{\sqrt{u\lambda} Q_0^2}, \qquad
    \psi = \sqrt{\frac{u}{\lambda Q_0^4}} \tilde{\psi} = \sqrt{\frac{u}{\lambda}} \frac{\rho}{\rho_0 Q_0^2}. \label{eq:rescaling2}
\end{align}
Generally the transverse interaction strength $\alpha_{j=0,1,2}$ could vary with space and time, i.e., $\alpha_j=\alpha_j(\mathbf{r},t)$. Equation \ref{eq:T-PFC} can be written as
\begin{equation}
    \frac{\partial \psi}{\partial t} = -\bm{\nabla} \cdot \mathbf{J} = -\bm{\nabla} \cdot \left ( \mathbf{J}_{\rm T} + \mathbf{J}_{\rm PFC} \right ),
\end{equation}
where the PFC flux $\mathbf{J}_{\rm PFC} = -\bm{\nabla} \mu_{\rm PFC} = -\bm{\nabla} \delta F_{\rm PFC}/\delta\psi$ contributed by longitudinal interactions described in original PFC, and the flux generated by transverse interaction is given by (with $i,j=x,y$ here)
\begin{equation}
    J_{\mathrm{T},i} = -\psi \left ( \alpha_0 + \alpha_1 \bm{\nabla}^2 + \alpha_2 \bm{\nabla}^4 \right ) \epsilon_{ij} \partial_j \psi,
\end{equation}
that is,
\begin{equation}
    \mathbf{J}_{\rm T} = -\psi \left ( \alpha_0 + \alpha_1 \bm{\nabla}^2 + \alpha_2 \bm{\nabla}^4 \right ) \partial_y \psi ~\hat{x} + \psi \left ( \alpha_0 + \alpha_1 \bm{\nabla}^2 + \alpha_2 \bm{\nabla}^4 \right ) \partial_x \psi ~\hat{y}.
\end{equation}

In the case of spatially constant coefficients $\alpha_j$, the T-PFC equation \ref{eq:T-PFC} is reduced to
\begin{equation}
    \frac{\partial \psi}{\partial t} = \left [ \left ( \bm{\nabla} \psi \right ) \times \bm{\nabla} \left ( \alpha_1 \bm{\nabla}^2 \psi + \alpha_2 \bm{\nabla}^4 \psi \right ) \right ]_z + \bm{\nabla}^2 \left [ -\epsilon \psi + \left ( \bm{\nabla}^2 + q_0^2 \right )^2 \psi - g\psi^2 + \psi^3 \right ].
    \label{eq:TPFC}
\end{equation}
The contribution from transverse interaction is presented as the first term of Eq.\ \ref{eq:TPFC}, which breaks the 2D parity symmetry (in terms of 2D parity inversion of $x \rightarrow -x$ or $y \rightarrow -y$), leading to 2D chirality. It is also antisymmetric with respect to the exchange of $x$ and $y$.

\section*{Model parameterization and numerical simulations}

\subsection*{Model parameters}
The model parameters in our T-PFC modeling are summarized in Table \ref{table:parameters}, including the corresponding physical interpretations and typical values used in the simulations. All these parameters have been rescaled to be dimensionless, as shown in the previous section (see also Eqs.~\ref{eq:rescaling1} and \ref{eq:rescaling2}). It would be important to match the model to real experimental systems, which is useful for model applications. Although in principle the PFC parameters could be parameterized via the fitting of liquid-state two- and three-point direct correlation functions (see Eq.~\ref{eq:coefficients_DDFT}) calculated from atomistic simulations (e.g., molecular dynamics) of real materials, the corresponding results and data are usually very limited or unknown, particularly for active or living systems. Here we apply an alternative approach of model parameterization developed for the PFC models of 2D materials \cite{TahaPRL17}, to match to length and energy density (or stress) scales of specific physical systems. To convert a dimensionless length scale $l$ to real units (with the corresponding dimensional quantities denoted by superscript ``$d$''), we identify the scaled length
\begin{equation}
    \gamma_l \equiv \frac{l^d}{l} = \frac{a_0^d}{a_0} = \frac{a_0^d}{4\pi/(\sqrt{3}\tilde{q}_0)}, \label{eq:gamma_l}
\end{equation}
where $a_0$ or $a_0^d$ is the lattice constant and $\tilde{q}_0$ is the selected wave number in the steady state, with $\tilde{q}_0 \approx q_0=1$. The value of $\gamma_l$ gives the physical length unit for the T-PFC model. (i) For living odd crystals of starfish embryos \cite{TanNature22}, $a_0^d \simeq 220 ~\mu\textrm{m}$, resulting in $\gamma_l \simeq 30.323 ~\mu\textrm{m}$ from Eq.~\ref{eq:gamma_l}. In our simulations conducted, a simulation size of $512 \times 512$ grid points corresponds to a physical size of $512 \Delta x^d \times 512 \Delta y^d = 1.22 \textrm{ cm} \times 1.22 \textrm{ cm}$, where $\Delta x^d = \Delta y^d = \gamma_l \Delta x$ with grid spacing $\Delta x = \Delta y = \pi/4$, and a large system of $2048 \times 2048$ grid size corresponds to $4.88 \textrm{ cm} \times 4.88 \textrm{ cm}$. (ii) For magnetic colloidal odd crystals \cite{BililignNatPhys22}, $a_0^d \simeq 2 ~\mu\textrm{m}$, yielding $\gamma_l \simeq 0.2757 ~\mu\textrm{m}$, such that a $512 \times 512$ simulated system corresponds to a size of $110.9 ~\mu\textrm{m} \times 110.9 ~\mu\textrm{m}$ and a $2048 \times 2048$ system corresponds to $443.5 ~\mu\textrm{m} \times 443.5 ~\mu\textrm{m}$. All these physical values are consistent with the experimental setups in those two odd crystalline systems.

\begin{table}\centering
\caption{Summary of model parameters in the dimensionless T-PFC equation, including their physical interpretations and typical values used in simulations.}
\label{table:parameters}
\begin{tabular}{clc}
\hline
Model parameter \\(dimensionless) & Physical interpretation & Values used \\
\hline
$\epsilon$ & control of transition between liquid (homogeneous) and solid states & 0.1 \\
$q_0$ & characteristic wave number of lattice structure & 1 \\
$g$ & control of cubic coupling in the PFC free energy & 0.5 \\
$\alpha_1$ & rescaled strength of transverse interactions & 0 -- 3 \\
$\alpha_2$ & rescaled strength of transverse interactions & 0 \\
$\alpha_0^0$ & magnitude of spatially varying transverse interaction strength $\alpha_0(\mathbf{r})$ & 1 \\
$\alpha_1^0$ & magnitude of spatially varying transverse interaction strength $\alpha_1(\mathbf{r})$ & 1 \\
$\bar{\psi}_0$ & average rescaled particle density & $-0.115$ to 0 \\
\hline
\end{tabular}
\end{table}

To determine the physical scale of 2D elastic moduli or stress in the T-PFC model (which is the same as the energy density scale of the crystalline bulk state), we use the scaling
\begin{equation}
    \gamma_\sigma \equiv \frac{\sigma_{ij}^d}{\sigma_{ij}} = \frac{B^d}{B} = \frac{\mu^d}{\mu}, \label{eq:gamma_sigma}
\end{equation}
where $\sigma_{ij}$ or $\sigma_{ij}^d$ is the stress tensor, $B$ or $B^d$ is the 2D bulk modulus, and $\mu$ or $\mu^d$ is the 2D shear modulus. In this T-PFC model, we have $B = 2\Gamma \tilde{q}_0^4 \simeq 6A_0^2 q_0^4$ and $\mu = \Gamma \tilde{q}_0^2 (3\tilde{q}_0^2 - 2q_0^2) \simeq 3A_0^2 q_0^4$ when $\tilde{q}_0 \approx q_0 =1$ (see Eq.~\ref{eq:elastic_moduli} below from elasticity analysis), where $\Gamma \simeq 3A_0^2$ and $A_0$ is the one-mode amplitude solution given in Eq. 7 of the main text (also Eq.~\ref{eq:one-mode} below). The $\gamma_\sigma$ value calculated from Eq.~\ref{eq:gamma_sigma} gives the real physical unit of stress or elastic moduli in the model. The estimated values of $B^d$ and $\mu^d$ are available only for starfish-embryo living crystals \cite{TanNature22}, giving $B^d \approx 7.7\times 10^{-7}$ N/m and $\mu^d \approx 3.85 \times 10^{-7}$ N/m. This leads to the same ratio of $B^d/\mu^d = B/\mu =2$ when $\tilde{q}_0=q_0$, also demonstrating the consistency between properties of T-PFC modeling and real experimental systems. Equation \ref{eq:gamma_sigma} then gives the real unit of $\gamma_\sigma \simeq 1.3 \times 10^{-5}$ N/m.

The dimensional model parameters for the transverse interaction coefficients, $\mathcal{A}_0$, $\mathcal{A}_1$, and $\mathcal{A}_2$ (before rescaling), can be estimated from Eqs.~\ref{eq:A0}--\ref{eq:A2}, or more specifically, Eqs.~\ref{eq:A0_hydro}--\ref{eq:A2_hydro} for living crystals of starfish embryos. For the transverse force given in Eq.~\ref{eq:starfish} between embryos \cite{TanNature22}, we have $r_0=110 ~\mu\textrm{m}$, $D_{\rm T} \beta f_0^\perp = r_0 (\omega_i+\omega_j) f_0 \simeq 3.96 \times 10^{-7}$ m/s where $f_0=0.06$ and $\omega_i \simeq \omega_j =0.03$ rad/s for any particle $i$ or $j$, and $r_c=r_0$ \cite{Choi24}. The physical parameter values of $\mathcal{A}_j$ coefficients corresponding to this living system are then given by
\begin{equation}
    D_{\rm T}\mathcal{A}_0 = 8.46 \times 10^{-18} \textrm{ m$^4$/s}, \qquad D_{\rm T}\mathcal{A}_1 = 6.81 \times 10^{-26} \textrm{ m$^6$/s}, \qquad D_{\rm T}\mathcal{A}_2 = 1.91 \times 10^{-34} \textrm{ m$^8$/s},
\end{equation}
with $D_{\rm T}$ the translational diffusion coefficient of starfish embryos, or if written in terms of the same unit,
\begin{equation}
    D_{\rm T}\mathcal{A}_0 = 8.46 \times 10^{-18} \textrm{ m$^4$/s}, \qquad D_{\rm T}\mathcal{A}_1/r_0^2 = 5.63 \times 10^{-18} \textrm{ m$^4$/s}, \qquad D_{\rm T}\mathcal{A}_2/r_0^4 = 1.31 \times 10^{-18} \textrm{ m$^4$/s}.
\end{equation}

\subsection*{Numerical simulations}
In our numerical simulations of the T-PFC model, three types of initial conditions have been used. (i) The first one is the setup of a single or multiple orderly placed circular crystalline grains coexisting with the surrounding liquid or homogeneous medium. This corresponds to the setups shown in Figs.~1 and 2 (for a single circular crystallite), Fig.~3 (two grains), and Fig.~5 (either one or eight equally spaced grains) of the main text, where results from one simulation run are shown. For most of them, the initial grains are single-crystalline, other than Fig. 1 C-E where individual dislocations or a grain boundary separating misoriented crystalline regions are initialized. (ii) The second type of initial condition is set up as multiple randomly distributed nuclei, with results presented in Fig. 4 of the main text. The nuclei are single-crystalline, all of the same initial radius (of 10 grid points each) but of different initial lattice orientation of each grain and spatial location. (iii) The third type is of random initial conditions that give spatially homogeneous states of the density field, with simulation outcomes also shown in Fig. 4. Results given in Fig. 4 D-H of the main text have been averaged over 20 independent runs, each for a system size of $2048 \times 2048$ grid points (while our tests for larger systems of size $4096 \times 4096$ are conducted for 5 runs). For the type (ii) initial setup of multiple crystalline nuclei, each run corresponds to different randomly distributed grain orientations and locations, while for the type (iii) setup, each run uses a different random number seed to generate a different spatial distribution of the homogeneous initial state.

Since we use a pseudospectral method to numerically solve the T-PFC model equation, the dependence of computational cost on the system size $N$ ($=L_x \times L_y$) is mainly determined by the algorithm of Fast Fourier Transform (FFT), for which we use the FFTW library with an algorithm that scales as $\mathcal{O}(N \log N)$. For the example of one simulation run of a $2048 \times 2048$ system up to time $t=10^5$ (as in Fig. 4 of the main text), it costs about 24 hours if running on 8 AMD 7313 CPUs in a cluster node with the use of an OpenMP code.

\section*{Model analysis}

\subsection*{One-mode approximation}

Consider the one-mode approximation for a crystalline state governed by Eq.\ \ref{eq:TPFC} with constant transverse interaction strength, i.e.,
\begin{equation}
    \psi = \bar{\psi}_0 + \sum_j A_j^0 e^{i \mathbf{q}_j^0 \cdot \mathbf{r}} + \textrm{c.c.},
    \label{eq:psi_onemode}
\end{equation}
where $\bar{\psi}_0$ is the average density variation and for a 2D hexagonal lattice structure with the steady-state selected wave number $\tilde{q}_0$ ($\sim q_0$), the basic wave vectors $\mathbf{q}_j^0$ ($j=1,2,3$) are given by
\begin{equation}
  \mathbf{q}_1^0 = \tilde{q}_0 \left ( -\frac{\sqrt{3}}{2}\hat{x} - \frac{1}{2}\hat{y} \right ),
  \qquad \mathbf{q}_2^0 = \tilde{q}_0 \hat{y}, \qquad
  \mathbf{q}_3^0 = \tilde{q}_0 \left ( \frac{\sqrt{3}}{2}\hat{x} - \frac{1}{2}\hat{y} \right ).
  \label{eq:q_j^0}
\end{equation}
In the steady state of a perfect crystal the amplitudes $A_j^0$ are constant, and thus the corresponding one-mode expansion $(\partial_x \psi) (\partial_y \bm{\nabla}^2 \psi) = (\partial_y \psi) (\partial_x \bm{\nabla}^2 \psi) = -\tilde{q}_0^2 \big (\sum_j iq_{jx}^0 A_j^0 e^{i \mathbf{q}_j^0 \cdot \mathbf{r}} + \textrm{c.c.} \big ) \big (\sum_k iq_{ky}^0 A_k^0 e^{i \mathbf{q}_k^0 \cdot \mathbf{r}} + \textrm{c.c.} \big )$ and $(\partial_x \psi) (\partial_y \bm{\nabla}^4 \psi) = (\partial_y \psi) (\partial_x \bm{\nabla}^4 \psi) = \tilde{q}_0^4 \big (\sum_j iq_{jx}^0 A_j^0 e^{i \mathbf{q}_j^0 \cdot \mathbf{r}} + \textrm{c.c.} \big ) \big (\sum_k iq_{ky}^0 A_k^0 e^{i \mathbf{q}_k^0 \cdot \mathbf{r}} + \textrm{c.c.} \big )$, which leads to no contribution from the transverse terms, i.e., $[ ( \bm{\nabla} \psi ) \times \bm{\nabla} \left ( \alpha_1 \bm{\nabla}^2 \psi + \alpha_2 \bm{\nabla}^4 \psi \right ) ]_z = 0$ in one-mode approximation for a perfect crystalline state (it would be still nonzero in the presence of interface or defects when $A_j^0$ are not constant in space). This is consistent with the microscopic picture that inside a perfect crystal transverse interactions between atoms are balanced and thus do not affect the structure stability (at least to the lowest order), and the imbalance of transverse interactions would occur around the interfaces/boundaries or defects, as shown in the schematics of Fig.~1 of the main text. Also only transverse interactions (without longitudinal ones) cannot maintain the crystal structure, as can be seen in particle-based simulations \cite{Choi24} and also from Eqs.~\ref{eq:TPFC} and \ref{eq:psi_onemode} that $d\bar{\psi}_0/dt=0$ ($\bar{\psi}_0 = \textrm{const.}$) and $dA_j^0/dt = -D_\textrm{T} \tilde{q}_0^2 A_j^0$ (with a standard diffusion term $D_{\rm T} \nabla^2 \rho$ but without the PFC terms), giving $A_j^0 \rightarrow 0$ at large time $t$, i.e., the dissolving of crystalline state.

Thus the one-mode solution for a perfect crystalline state of 2D hexagonal symmetry is the same as original PFC, with
\begin{equation}
    A_j^0 \equiv A_0 = \frac{1}{15} \left \{ g - 3\bar{\psi}_0 \pm \sqrt{(g-3\bar{\psi}_0)^2 - 15\left [ -\epsilon + (\tilde{q}_0^2 - q_0^2)^2 + 3\bar{\psi}_0^2 - 2g\bar{\psi}_0 \right ]} \right \}.
    \label{eq:one-mode}
\end{equation}
When $g-3\bar{\psi}_0 > 0$, ``$+$'' sign is chosen with $A_0 > 0$, corresponding to a triangular lattice, while if $g-3\bar{\psi}_0 < 0$, ``$-$'' sign is chosen with $A_0 < 0$, corresponding to a honeycomb lattice (i.e., inverse triangular, like graphene). The corresponding one-mode phase diagram is also the same as that of original PFC \cite{Provatas10}.

\subsection*{Amplitude equations for T-PFC}

It would be useful to derive the amplitude equations for this T-PFC model, from which some analytic results can be better obtained, in addition to simulations at larger mesoscopic scales. In the amplitude expansion, the mesoscopic ``slow'' scales $(X = \varepsilon x, Y = \varepsilon y, T = \varepsilon t)$ for $\psi_0$ and complex amplitudes $A_j$, with $\varepsilon$ a dimensionless small parameter, are separated from the microscopic ``fast'' crystalline scales $(x,y,t)$, such that
\begin{equation}
    \psi = \psi_0(X,Y,T) + \sum_j A_j(X,Y,T) e^{i\mathbf{q}_j^0 \cdot \mathbf{r}} + \textrm{c.c.},
\end{equation}
with basic wave vectors $\mathbf{q}_j^0$ given in Eq.\ \ref{eq:q_j^0}. Applying the procedure of amplitude formulation for the PFC model \cite{GoldenfeldPRE05,HuangPRE10} to Eq.\ \ref{eq:TPFC}, we obtain the amplitude equations  for T-PFC (back to the original scales of $(x,y,t)$, with constant $\alpha_j$ coefficients) as
\begin{align}
    \frac{\partial \psi_0}{\partial t} =&~ \left [ \left ( \bm{\nabla} \psi_0 \right ) \times \bm{\nabla} \left ( \alpha_1 \bm{\nabla}^2 \psi_0 + \alpha_2 \bm{\nabla}^4 \psi_0 \right ) \right ]_z \nonumber\\
    &~ + \sum_j \left \{ \left [ \left ( \bm{\nabla} + i\mathbf{q}_j^0 \right ) A_j \right ] \times \left ( \bm{\nabla} - i\mathbf{q}_j^0 \right ) \right \}_z \left ( \mathcal{G}_j^* - \tilde{q}_0^2 \right ) \left [ \alpha_1 + \alpha_2 \left ( \mathcal{G}_j^* - \tilde{q}_0^2 \right ) \right ] A_j^* + \textrm{c.c.} \nonumber\\
    &~ + \bm{\nabla}^2 \left [ \left ( -\epsilon + q_0^4 \right ) \psi_0 - g\psi_0^2 + \psi_0^3 + (6\psi_0 - 2g) \sum_j |A_j|^2 + 6 \left ( A_1A_2A_3 + \textrm{c.c.} \right ) \right ], \label{eq:psi0}\\
    \frac{\partial A_j}{\partial t} =&~ \left [ \left ( \bm{\nabla} \psi_0 \right ) \times \left ( \bm{\nabla} + i\mathbf{q}_j^0 \right ) \right ]_z \left ( \mathcal{G}_j - \tilde{q}_0^2 \right ) \left [ \alpha_1 + \alpha_2 \left ( \mathcal{G}_j - \tilde{q}_0^2 \right ) \right ] A_j
    + \left \{ \left [ \left ( \bm{\nabla} + i\mathbf{q}_j^0 \right ) A_j \right ] \times \bm{\nabla} \left ( \alpha_1 \bm{\nabla}^2 \psi_0 + \alpha_2 \bm{\nabla}^4 \psi_0 \right ) \right \}_z \nonumber\\
    &~ + \sum_{l \neq k \neq j} \left \{ \left [ \left ( \bm{\nabla} - i\mathbf{q}_l^0 \right ) A_l^* \right ] \times \left ( \bm{\nabla} - i\mathbf{q}_k^0 \right ) \right \}_z \left ( \mathcal{G}_k^* - \tilde{q}_0^2 \right ) \left [ \alpha_1 + \alpha_2 \left ( \mathcal{G}_k^* - \tilde{q}_0^2 \right ) \right ] A_k^* \nonumber\\
    &~ - \tilde{q}_0^2 \left [ \left ( -\epsilon + 3\psi_0^2 - 2g\psi_0 \right ) A_j + \left ( \mathcal{G}_j - \tilde{q}_0^2 + q_0^2 \right )^2 A_j + (6\psi_0 - 2g) \prod_{k \neq j} A_k^* + 3A_j \left ( |A_j|^2 + 2\sum_{k \neq j} |A_k|^2 \right ) \right ], \label{eq:Aj}
\end{align}
where for the PFC terms (i.e., the last terms of Eqs.~\ref{eq:psi0} and \ref{eq:Aj}) we have used the long wavelength approximation as before \cite{HuangPRE10,HuangPRE13}, and
\begin{equation}
    \mathcal{G}_j = \mathcal{L}_j + \tilde{q}_0^2 = \bm{\nabla}^2 + 2i\mathbf{q}_j^0 \cdot \bm{\nabla}.
\end{equation}
The last terms of Eqs.~\ref{eq:psi0} and \ref{eq:Aj} can be rewritten as $\bm{\nabla}^2 \delta \mathcal{F}_{\rm PFC} / \delta \psi_0$ and $-\tilde{q}_0^2 \delta \mathcal{F}_{\rm PFC} / \delta A_j^*$, respectively, with $\mathcal{F}_{\rm PFC}[A_j,\psi_0]$ the effective PFC free energy functional given by \cite{HuangPRE13}
\begin{align}
    \mathcal{F}_{\rm PFC} =& \int d\mathbf{r} \left [ \left ( -\epsilon + 3\psi_0^2 - 2g\psi_0 \right ) \sum_j |A_j|^2 + \sum_j \left | \left ( \mathcal{G}_j - \tilde{q}_0^2 + q_0^2 \right ) A_j \right |^2 + (6\psi_0 - 2g) \left ( \prod_j A_j + \textrm{c.c.} \right ) \right. \nonumber\\
    & \left. + \frac{3}{2} \sum_j |A_j|^4 + 6\sum_{j<k} |A_j|^2 |A_k|^2 + \frac{1}{2} \left (-\epsilon + q_0^4 \right ) \psi_0^2 - \frac{1}{3} g\psi_0^3 + \frac{1}{4} \psi_0^4 \right ].
\end{align}

\subsection*{Elastodynamics}

It is important to examine whether this new T-PFC model with transverse interaction incorporates the property of odd elasticity. In the limit of small deformation with a displacement field $\mathbf{u}$ varying slowly at long wavelength, when $\mathbf{r} \rightarrow \mathbf{r} + \mathbf{u}$ we have amplitudes $A_j \simeq A_j^0 \exp(-i\mathbf{q}_j^0 \cdot \mathbf{u})$, where in the steady state $\partial A_j^0 / \partial t = \partial \psi_0 / \partial t = 0$ and $\bm{\nabla} A_j^0 = \bm{\nabla} \psi_0 = 0$. To the first order of displacements $\mathcal{O}(\mathbf{u})$, the amplitude equations \ref{eq:psi0} and \ref{eq:Aj} give
\begin{equation}
    \sum_j |A_j^0|^2 \left ( \mathbf{q}_j^0 \times \bm{\nabla} \right )_z \left [ \alpha_1 + \alpha_2 \left ( \mathcal{G}_j^* - 2\tilde{q}_0^2 \right ) \right ] \mathcal{G}_j^* \left ( \mathbf{q}_j^0 \cdot \mathbf{u} \right ) = 0,
    \label{eq:u_psi0}
\end{equation}
and
\begin{align}
    -iA_j^0 \frac{\partial}{\partial t} \left ( \mathbf{q}_j^0 \cdot \mathbf{u} \right ) = & \sum_{l \neq k \neq j} {A_l^0}^* {A_k^0}^* \left [ \mathbf{q}_l^0 \times \left (\bm{\nabla} - i\mathbf{q}_k^0 \right ) \right ]_z \left [ \alpha_1 + \alpha_2 \left ( \mathcal{G}_k^* - 2\tilde{q}_0^2 \right ) \right ] \mathcal{G}_k^* \left ( \mathbf{q}_k^0 \cdot \mathbf{u} \right ) \nonumber\\
    & + i\tilde{q}_0^2 A_j^0 \left [ \mathcal{G}_j^2 - 2\left ( \tilde{q}_0^2 - q_0^2 \right ) \mathcal{G}_j \right ] \left ( \mathbf{q}_j^0 \cdot \mathbf{u} \right ) \nonumber\\
    & - \tilde{q}_0^2 \left \{ \left [ -\epsilon + \left ( \tilde{q}_0^2 - q_0^2 \right )^2 + 3\psi_0^2 - 2g\psi_0 \right ] A_j^0 + (6\psi_0 - 2g) \prod_{k \neq j} A{_k^0}^* + 3A_j^0 \left ( |A_j^0|^2 + 2\sum_{k \neq j} |A_k^0|^2 \right ) \right \}. \label{eq:u_Aj}
\end{align}
If assuming in the steady state of a crystalline lattice $A_j^0 \simeq A_0$ (i.e., the one-mode solution given in Eq.\ \ref{eq:one-mode} so that the last line of Eq.\ \ref{eq:u_Aj} is approximately equal to zero), we can obtain the following conditions at $\mathcal{O}(\mathbf{u})$ by separating the real and imaginary parts of Eq.\ \ref{eq:u_psi0},
\begin{gather}
    \left [ \alpha_1 + \alpha_2 \left ( \bm{\nabla}^2 - 3\tilde{q}_0^2 \right ) \right ] \bm{\nabla}^2 \left ( \partial_x u_y - \partial_y u_x \right ) = 0, \label{eq:condR_psi0}\\
    \left [ \alpha_1 + 2\alpha_2 \left ( \bm{\nabla}^2 - \tilde{q}_0^2 \right ) \right ] \left ( \partial_x^2 u_x - \partial_y^2 u_x - 2\partial_x \partial_y u_y \right ) = 0. \label{eq:condI_psi0}
\end{gather}
Similarly, the real part of Eq.\ \ref{eq:u_Aj} leads to
\begin{align}
    & A_0 \sum_{l \neq k \neq j} \Bigl \{ \left ( \mathbf{q}_l^0 \times \bm{\nabla} \right )_z \left [ \left [ \alpha_1 + \alpha_2 \left ( \bm{\nabla}^2 - 2\tilde{q}_0^2 \right ) \right ] \bm{\nabla}^2 - 4 \alpha_2 \left ( \mathbf{q}_k^0 \cdot \bm{\nabla} \right )^2 \right ] \nonumber\\
    & \qquad - 2\left ( \mathbf{q}_l^0 \times \mathbf{q}_k^0 \right )_z \left ( \mathbf{q}_k^0 \cdot \bm{\nabla} \right ) \left [ \alpha_1 + 2\alpha_2 \left ( \bm{\nabla}^2 - \tilde{q}_0^2 \right ) \right ] \Bigr \} \left ( \mathbf{q}_k^0 \cdot \mathbf{u} \right )
    -4\tilde{q}_0^2 \left ( \mathbf{q}_j^0 \cdot \bm{\nabla} \right ) \left [ \bm{\nabla}^2 - \left ( \tilde{q}_0^2 - q_0^2 \right ) \right ] \left ( \mathbf{q}_j^0 \cdot \mathbf{u} \right ) = 0, \label{eq:u_real_Aj}
\end{align}
and for the imaginary part,
\begin{align}
    \frac{\partial}{\partial t} \left ( \mathbf{q}_j^0 \cdot \mathbf{u} \right ) = &~ A_0 \sum_{l \neq k \neq j} \Bigl \{ 2\left ( \mathbf{q}_l^0 \times \bm{\nabla} \right )_z \left ( \mathbf{q}_k^0 \cdot \bm{\nabla} \right ) \left [ \alpha_1 + 2\alpha_2 \left ( \bm{\nabla}^2 - \tilde{q}_0^2 \right ) \right ] \nonumber\\
    & \qquad + \left ( \mathbf{q}_l^0 \times \mathbf{q}_k^0 \right )_z \left [ \left [ \alpha_1 + \alpha_2 \left ( \bm{\nabla}^2 - 2\tilde{q}_0^2 \right ) \right ] \bm{\nabla}^2 - 4 \alpha_2 \left ( \mathbf{q}_k^0 \cdot \bm{\nabla} \right )^2 \right ] \Bigr \} \left ( \mathbf{q}_k^0 \cdot \mathbf{u} \right ) \nonumber\\
    &~ - \tilde{q}_0^2 \left [ \bm{\nabla}^4 - 2\left ( \tilde{q}_0^2 - q_0^2 \right ) \bm{\nabla}^2 - 4 \left ( \mathbf{q}_j^0 \cdot \bm{\nabla} \right )^2 \right ] \left ( \mathbf{q}_j^0 \cdot \mathbf{u} \right ). \label{eq:u_img_Aj}
\end{align}

At $j=2$ Eq.\ \ref{eq:u_real_Aj} gives
\begin{align}
    & -2A_0 \left [ \alpha_1 + \alpha_2 \left ( \bm{\nabla}^2 - 3\tilde{q}_0^2 \right ) \right ] \bm{\nabla}^2 \partial_y u_x + \alpha_2 A_0 \tilde{q}_0^2 \partial_x \left [ 6 \partial_x \partial_y u_x + \left ( \partial_x^2 + 3\partial_y^2 \right ) u_y \right ] \nonumber\\
    & + \frac{3}{2} A_0 \tilde{q}_0^2 \left [ \alpha_1 + 2\alpha_2 \left ( \bm{\nabla}^2 - \tilde{q}_0^2 \right ) \right ] \left ( \partial_x u_y + \partial_y u_x \right ) - 4\tilde{q}_0^2 \left [ \bm{\nabla}^2 - \left ( \tilde{q}_0^2 - q_0^2 \right ) \right ] \partial_y u_y =0, \label{eq:condR_A2}
\end{align}
where Eq.\ \ref{eq:condR_psi0} has been used. Adding expressions of Eq.\ \ref{eq:u_real_Aj} at $j=1$ and $j=3$ and also with Eq.\ \ref{eq:condR_A2} yields
\begin{equation}
    \left [ \bm{\nabla}^2 - \left ( \tilde{q}_0^2 - q_0^2 \right ) \right ] \left ( \partial_x u_x + \partial_y u_y \right ) = 0. \label{eq:condR_A13_2}
\end{equation}
Subtracting between expressions of Eq.\ \ref{eq:u_real_Aj} at $j=1$ and $j=3$ leads to
\begin{align}
    & -A_0 \left [ \alpha_1 + \alpha_2 \left ( \bm{\nabla}^2 - 3\tilde{q}_0^2 \right ) \right ] \bm{\nabla}^2 \left ( \partial_x u_x - \partial_y u_y \right ) + \alpha_2 A_0 \tilde{q}_0^2 \left [ 2\partial_x^3 u_x + 3 \left ( \partial_x^2 - \partial_y^2 \right ) \partial_y u_y \right ] \nonumber\\
    & + \frac{3}{2} A_0 \tilde{q}_0^2 \left [ \alpha_1 + 2\alpha_2 \left ( \bm{\nabla}^2 - \tilde{q}_0^2 \right ) \right ] \left ( \partial_x u_x - \partial_y u_y \right ) - 2\tilde{q}_0^2 \left [ \bm{\nabla}^2 - \left ( \tilde{q}_0^2 - q_0^2 \right ) \right ] \left ( \partial_x u_y + \partial_y u_x \right ) = 0.
\end{align}

From expressions of Eq.\ \ref{eq:u_img_Aj} at $j=1$ and $j=3$ we get
\begin{align}
    \frac{\partial u_x}{\partial t} =&~ A_0 \tilde{q}_0^2 \Bigl \{ - \left [ \alpha_1 + 2\alpha_2 \left ( \bm{\nabla}^2 - \tilde{q}_0^2 \right ) \right ] \left [ \partial_x \partial_y u_x + \left ( \partial_x^2 + 2\partial_y^2 \right ) u_y \right ] - \frac{3}{2} \left [ \alpha_1 + \alpha_2 \left ( \bm{\nabla}^2 - 3\tilde{q}_0^2 \right ) \right ] \bm{\nabla}^2 u_y \nonumber\\
    &~ \qquad + 3\alpha_2 \tilde{q}_0^2 \partial_y \left ( \partial_x u_x + \partial_y u_y \right ) \Bigr \}
    - \tilde{q}_0^2 \left [ \bm{\nabla}^4 - 2\left ( \tilde{q}_0^2 - q_0^2 \right ) \bm{\nabla}^2 - \tilde{q}_0^2 \left ( 3\partial_x^2 + \partial_y^2 \right ) \right ] u_x + 2\tilde{q}_0^4 \partial_x \partial_y u_y, \label{eq:ux}\\
    \frac{\partial u_y}{\partial t} =&~ A_0 \tilde{q}_0^2 \Bigl \{ \left [ \alpha_1 + 2\alpha_2 \left ( \bm{\nabla}^2 - \tilde{q}_0^2 \right ) \right ] \left ( 3\partial_x^2 u_x - \partial_x \partial_y u_y \right ) + \frac{3}{2} \left [ \alpha_1 + \alpha_2 \left ( \bm{\nabla}^2 - 3\tilde{q}_0^2 \right ) \right ] \bm{\nabla}^2 u_x \nonumber\\
    &~ \qquad - 3\alpha_2 \tilde{q}_0^2 \partial_x \left ( \partial_x u_x + \partial_y u_y \right ) \Bigr \}
    + 6\tilde{q}_0^4 \partial_x \partial_y u_x - \tilde{q}_0^2 \left [ \bm{\nabla}^4 - 2\left ( \tilde{q}_0^2 - q_0^2 \right ) \bm{\nabla}^2 - \tilde{q}_0^2 \left ( 3\partial_x^2 + \partial_y^2 \right ) \right ] u_y, \label{eq:uy1}
\end{align}
while at $j=2$ we have
\begin{align}
    \frac{\partial u_y}{\partial t} =&~ A_0 \tilde{q}_0^2 \Bigl \{ \frac{1}{2} \left [ \alpha_1 + 2\alpha_2 \left ( \bm{\nabla}^2 - \tilde{q}_0^2 \right ) \right ] \left ( 3\bm{\nabla}^2 u_x + 4\partial_x \partial_y u_y \right ) + \frac{3}{2} \left [ \alpha_1 + \alpha_2 \left ( \bm{\nabla}^2 - 3\tilde{q}_0^2 \right ) \right ] \bm{\nabla}^2 u_x \nonumber\\
    &~ \qquad - 3\alpha_2 \tilde{q}_0^2 \partial_x \left ( \partial_x u_x + \partial_y u_y \right ) \Bigr \}
    - \tilde{q}_0^2 \left [ \bm{\nabla}^4 - 2\left ( \tilde{q}_0^2 - q_0^2 \right ) \bm{\nabla}^2 - 4\tilde{q}_0^2 \partial_y^2 \right ] u_y. \label{eq:uy2}
\end{align}
Using Eq.\ \ref{eq:condI_psi0} it can be shown that the first term of Eqs.~\ref{eq:uy1} and \ref{eq:uy2} is equal to each other at $\mathcal{O}(\mathbf{u})$, but the rest are different. These two equations then give the following condition at the first order of displacements
\begin{equation}
    2\partial_x \partial_y u_x + \left ( \partial_x^2 - \partial_y^2 \right ) u_y = 0.
\end{equation}
To remove the above ambiguity for the dynamical equation of $u_y$, we note that at the limit of $\alpha_1=\alpha_2=0$ one can return to the result of original PFC for regular, even elasticity (with the major symmetry of elastic constants $C_{ijkl} = C_{klij}$). We thus need to choose a form consistent with the symmetric results at that limit, which is obtained via multiplying Eq.\ \ref{eq:uy1} by $1/3$ and Eq.\ \ref{eq:uy2} by $2/3$ and adding them together, as done for original PFC. This gives
\begin{align}
    \frac{\partial u_y}{\partial t} =&~ A_0 \tilde{q}_0^2 \Bigl \{ \left [ \alpha_1 + 2\alpha_2 \left ( \bm{\nabla}^2 - \tilde{q}_0^2 \right ) \right ] \left [ \left ( 2\partial_x^2 + \partial_y^2 \right ) u_x + \partial_x \partial_y u_y \right ] + \frac{3}{2} \left [ \alpha_1 + \alpha_2 \left ( \bm{\nabla}^2 - 3\tilde{q}_0^2 \right ) \right ] \bm{\nabla}^2 u_x \nonumber\\
    &~ \qquad - 3\alpha_2 \tilde{q}_0^2 \partial_x \left ( \partial_x u_x + \partial_y u_y \right ) \Bigr \}
    + 2\tilde{q}_0^4 \partial_x \partial_y u_x - \tilde{q}_0^2 \left [ \bm{\nabla}^4 - 2\left ( \tilde{q}_0^2 - q_0^2 \right ) \bm{\nabla}^2 - \tilde{q}_0^2 \left ( \partial_x^2 + 3\partial_y^2 \right ) \right ] u_y. \label{eq:uy}
\end{align}
Equations \ref{eq:ux} and \ref{eq:uy} constitute the linear elastodynamical equations for the displacement field, in the overdamped limit of a crystalline state incorporating both longitudinal and transverse interactions. It can be seen that as long as $\alpha_1 \neq 0$ and/or $\alpha_2 \neq 0$, these elastodynamical equations are no longer symmetric with respect to $x \leftrightarrow y$, and the corresponding dynamical matrix becomes non-Hermitian, indicating the property of nonreciprocity of the system. This is caused by the $x \leftrightarrow y$ antisymmetry of the $\alpha_1$ and $\alpha_2$ terms, as originating from the corresponding antisymmetric property in the T-PFC equation \ref{eq:TPFC}. The rest of terms representing effects of longitudinal interactions (i.e., terms from original PFC) are symmetric with respect to $x \leftrightarrow y$.

To be specific, consider the general form of elastodynamical equation for the displacement field $u_j$ ($j=x,y$)
\begin{equation}
    \Gamma \frac{\partial u_j}{\partial t} = \partial_i \sigma_{ij} = C_{ijkl} \partial_i \partial_k u_l,
    \label{eq:uj_elastodynamic}
\end{equation}
which is the overdamped version of Cauchy's equation of motion. Here $\Gamma$ is the drag or friction coefficient and in linear elasticity the stress tensor $\sigma_{ij} = C_{ijkl} \partial_k u_l$, with elasticity tensor $C_{ijkl}$. Matching Eqs.~\ref{eq:ux} and \ref{eq:uy} with the form of Eq.\ \ref{eq:uj_elastodynamic} and to the lowest order neglecting the fourth-order gradients of $u_x$ and $u_y$, we get
\begin{align}
    & C_{1111} = C_{2222} = \Gamma \tilde{q}_0^2 \left ( 5\tilde{q}_0^2 - 2q_0^2 \right ), \qquad
    C_{1122} = C_{2211} = \Gamma \tilde{q}_0^2 \left ( 2q_0^2 - \tilde{q}_0^2 \right ), \nonumber\\
    & C_{1212} = C_{2121} = C_{1221} = C_{2112} = \Gamma \tilde{q}_0^2 \left ( 3\tilde{q}_0^2 - 2q_0^2 \right ),
\end{align}
which are of the same form as the original PFC with even elasticity (for which $\tilde{q}_0^2 = q_0^2$ and $\Gamma = 3A_0^2$; see Ref.~\cite{ElderPRE10}). Effects of transverse interaction with nonzero $\alpha_1$ or $\alpha_2$ terms enter into the property of elasticity via extra elastic constants $C_{ijkk}$ ($i \neq j$) and $C_{iikl}$ ($k \neq l$). Since nonzero internal torques can be induced by transverse interaction (leading to the lack of angular momentum conservation), we could no longer assume the left minor symmetry for these elastic constants, so that $C_{ijkk} \neq C_{jikk}$ (and thus $\sigma_{ij} \neq \sigma_{ji}$ with $i \neq j$). Since the system still maintains global rotational invariance as shown in the T-PFC model equation \ref{eq:TPFC} and there is no coupling to rotational degrees of freedom, as in Ref.~\cite{ScheibnerNatPhys20} deformation dependence of the elastic response with right mirror symmetry $C_{iikl} = C_{iilk}$ can be assumed. We then have
\begin{align}
    & C_{1112} = C_{1121} = -\frac{1}{2}\Gamma A_0 \tilde{q}_0^2 \left (5\alpha_1 - 13\tilde{q}_0^2\alpha_2 \right ), 
    & C_{2221} &= C_{2212} = \frac{1}{2}\Gamma A_0 \tilde{q}_0^2 \left (5\alpha_1 - 13\tilde{q}_0^2\alpha_2 \right ), \nonumber\\
    & C_{1211} = \frac{1}{2}\Gamma A_0 \tilde{q}_0^2 \left (7\alpha_1 - 23\tilde{q}_0^2\alpha_2 \right ), 
    & C_{2111} &= \frac{3}{2}\Gamma A_0 \tilde{q}_0^2 \left (\alpha_1 - \tilde{q}_0^2\alpha_2 \right ), \nonumber\\
    & C_{2122} = -\frac{1}{2}\Gamma A_0 \tilde{q}_0^2 \left (7\alpha_1 - 23\tilde{q}_0^2\alpha_2 \right ),
    & C_{1222} &= -\frac{3}{2}\Gamma A_0 \tilde{q}_0^2 \left (\alpha_1 - \tilde{q}_0^2\alpha_2 \right ),
\end{align}
which violate both the major symmetry and left minor symmetry, i.e., $C_{ijkk} \neq C_{kkij}$, $C_{iikl} \neq C_{klii}$, and $C_{ijkk} \neq C_{jikk}$. It is noted that these elastic constants are antisymmetric with respect to $1 \leftrightarrow 2$ (i.e., the exchange of $x$ and $y$), with $C_{iikl} = -C_{jjlk}$ and $C_{ijkk} = -C_{jill}$ ($i \neq j$, $k \neq l$), which can be attributed to the $x \leftrightarrow y$ antisymmetry of the contributions of transverse interaction as seen from the $\alpha_1$ and $\alpha_2$ terms in Eqs.~\ref{eq:TPFC}, \ref{eq:ux}, and \ref{eq:uy}. 

The lack of major symmetry indicates the incorporation of both even and odd elasticity, showing as
\begin{equation}
    C_{iikl} = C_{iikl}^{\rm (e)} + C_{iikl}^{\rm (o)}, \qquad
    C_{ijkk} = C_{ijkk}^{\rm (e)} + C_{ijkk}^{\rm (o)},
\end{equation}
containing regular, even elasticity components with major symmetry $C^{(\rm e)}_{ijkl} = C^{(\rm e)}_{klij}$ as well as odd elasticity components with antisymmetric elastic constants $C^{(\rm o)}_{ijkl} = - C^{(\rm o)}_{klij}$ as a result of transverse interaction. The explicit expressions are given by
\begin{align}
    C_{1112}^{\rm (e)} &= C_{1211}^{\rm (e)} = \frac{1}{2}\Gamma A_0 \tilde{q}_0^2 \left (\alpha_1 - 5\tilde{q}_0^2\alpha_2 \right ),
    & C_{1112}^{\rm (o)} &= -C_{1211}^{\rm (o)} = -3\Gamma A_0 \tilde{q}_0^2 \left (\alpha_1 - 3\tilde{q}_0^2\alpha_2 \right ), \nonumber\\
    C_{1121}^{\rm (e)} &= C_{2111}^{\rm (e)} = -\frac{1}{2}\Gamma A_0 \tilde{q}_0^2 \left (\alpha_1 - 5\tilde{q}_0^2\alpha_2 \right ),
    & C_{1121}^{\rm (o)} &= -C_{2111}^{\rm (o)} = -2\Gamma A_0 \tilde{q}_0^2 \left (\alpha_1 - 2\tilde{q}_0^2\alpha_2 \right ), \nonumber\\
    C_{1222}^{\rm (e)} &= C_{2212}^{\rm (e)} = \frac{1}{2}\Gamma A_0 \tilde{q}_0^2 \left (\alpha_1 - 5\tilde{q}_0^2\alpha_2 \right ),
    & C_{1222}^{\rm (o)} &= -C_{2212}^{\rm (o)} = -2\Gamma A_0 \tilde{q}_0^2 \left (\alpha_1 - 2\tilde{q}_0^2\alpha_2 \right ), \nonumber\\
    C_{2122}^{\rm (e)} &= C_{2221}^{\rm (e)} = -\frac{1}{2}\Gamma A_0 \tilde{q}_0^2 \left (\alpha_1 - 5\tilde{q}_0^2\alpha_2 \right ),
    & C_{2122}^{\rm (o)} &= -C_{2221}^{\rm (o)} = -3\Gamma A_0 \tilde{q}_0^2 \left (\alpha_1 - 3\tilde{q}_0^2\alpha_2 \right ).
\end{align}

If using the basis of dilation, rotation, and two independent shear modes introduced in Ref.~\cite{ScheibnerNatPhys20}, i.e.,
\begin{equation}
    \tau^0=
    \begin{pmatrix}
        1 & 0 \\
        0 & 1
    \end{pmatrix}, \qquad
    \tau^1=
    \begin{pmatrix}
        0 & -1 \\
        1 & 0
    \end{pmatrix}, \qquad
    \tau^2=
    \begin{pmatrix}
        1 & 0 \\
        0 & -1
    \end{pmatrix}, \qquad
    \tau^3=
    \begin{pmatrix}
        0 & 1 \\
        1 & 0
    \end{pmatrix},
\end{equation}
we can obtain the same form of the corresponding elastic tensor as in the continuum odd elasticity theory \cite{ScheibnerNatPhys20}
\begin{equation}
    C^{\alpha\beta} = \frac{1}{2} \tau_{ij}^\alpha C_{ijkl} \tau_{kl}^\beta = 2
    \begin{pmatrix}
        B & 0 & 0 & 0 \\
        A & 0 & 0 & 0 \\
        0 & 0 & \mu & K^{\rm o} \\
        0 & 0 & -K^{\rm o} & \mu
    \end{pmatrix},
\end{equation}
with the bulk modulus $B$, the shear modulus $\mu$, the odd modulus $A$ caused by nonzero internal torque (with the violation of left minor symmetry of elastic constants), and another odd modulus $K^{\rm o}$ representing the antisymmetric coupling of two independent shear modes. All of them can be expressed explicitly in terms of T-PFC parameters, i.e.,
\begin{gather}
    B = \frac{1}{2} \left ( C_{1111} + C_{1122} \right ) = 2\Gamma \tilde{q}_0^4, \qquad
    \mu = \frac{1}{2} \left ( C_{1111} - C_{1122} \right ) = \Gamma \tilde{q}_0^2 \left ( 3\tilde{q}_0^2 - 2q_0^2 \right ), \nonumber\\
    A = \frac{1}{2} \left ( C_{2111} - C_{1211} \right ) = -\Gamma A_0 \tilde{q}_0^2 \left (\alpha_1 - 5\tilde{q}_0^2\alpha_2 \right ), \qquad
    K^{\rm o} = C_{1112} = -\frac{1}{2} \left ( C_{2111} + C_{1211} \right ) = -\frac{1}{2}\Gamma A_0 \tilde{q}_0^2 \left (5\alpha_1 - 13\tilde{q}_0^2\alpha_2 \right ).
    \label{eq:elastic_moduli}
\end{gather}

The linear instability analysis can be conducted on the elastodynamical equations \ref{eq:ux} and \ref{eq:uy}, or equivalently Eq.\ \ref{eq:uj_elastodynamic}, giving a non-Hermitian dynamical matrix in terms of $C_{ijkl}$, as well as the same form of results for perturbation growth rate and exceptional point as those of continuum odd elasticity theory in Ref.~\cite{ScheibnerNatPhys20}, with details expressed by T-PFC parameters.

These analytic results (and the numerical results presented in the main text) indicate that the lowest-order $\alpha_1$ term in T-PFC is sufficient to model all the important effects of odd elasticity and transverse interaction. This can be also justified from symmetry consideration, i.e., rotational invariance and 2D chirality with the breaking of parity symmetry, of the $\alpha_1$ term in the T-PFC equation.

\section*{Additional discussion of simulation results}

\subsection*{Grain self-rotation}
Results from both T-PFC simulations and experimental data of starfish embryo living crystals, as presented in Fig. 1B of the main text, show a power law scaling of $\omega \sim N^{-1}$, where $\omega$ and $N$ are the self-rotation frequency and particle number of the odd crystallite, respectively. This can be understood by noting that the steady-state frequency $\omega = \tau_{\rm net}/\zeta_R$ in the overdamped limit, where $\zeta_R$ is the rotational friction coefficient and $\tau_{\rm net}$ is the self-generated torque on the surface of crystallite as a result of net surface transverse force $F_{\rm net}^\perp \propto 2\pi R \alpha_1$ (given that interparticle transverse forces cancel out in the bulk of a perfect lattice), i.e., $\tau_{\rm net} = \int_s R dF_{\rm net}^\perp \propto R^2\alpha_1$ for a surface of 2D grain with radius $R$. To estimate the $R$-dependence of $\zeta_R$ in 2D, consider a disk-shaped cluster of $N$ particles, with area $\pi R^2 \propto N$. Each particle $i$ inside has a velocity $v_i$ and a translational friction coefficient $\gamma_f$ (same for all particles), subjected to the friction force $f_i = -\gamma_f v_i$. In the case of translation of the cluster, $v_i \equiv v_0$ for all particles, and thus the total friction force of the cluster is $F_\gamma = \sum_{i=1}^N f_i = -N \gamma_f v_0 \propto -R^2 \gamma_f v_0$. This leads to $\zeta_T \propto \gamma_f R^2$ in 2D, where $\zeta_T$ is the translational friction coefficient of the whole cluster. For the case of cluster rotation as addressed here, consider a ring within the cluster with a distance $r$ from the center and a thickness $dr$. In the steady state with constant angular velocity $\omega_0$ of the rotating cluster, the velocity of a particle $i$ in this ring is $v_i = \omega_0 r$, so that the friction force for this ring of $dN \propto 2\pi r dr$ particles is given by $dF_\gamma = \gamma_f v_i dN \propto 2\pi \gamma_f \omega_0 r^2 dr$. Hence the total frictional torque corresponding to the rotation of the whole cluster is
\begin{equation}
    M_\gamma = \int_0^R rdF_\gamma \propto -\int_0^R 2\pi \gamma_f \omega_0 r^3 dr = -\frac{\pi}{2} \gamma_f \omega_0 R^4,
\end{equation}
leading to a scaling of the 2D rotational friction coefficient $\zeta_R \propto \gamma_f R^4$. Substituting this into the above overdamped result for the self-rotating grain gives
\begin{equation}
    \omega = \tau_{\rm net}/\zeta_R \propto (R^2\alpha_1) / R^4 = \alpha_1 / R^2 \propto \alpha_1 / N,
    \label{eq:omega}
\end{equation}
consistent with the scaling relation found in numerical simulations of T-PFC and the fitting to experimental data.

\subsection*{Surface cusp instability}

The onset of surface cusp instability of self-rotating odd crystallites is found to obey a power-law scaling of $R_c \sim |\alpha_1|^{-2.5}$ from T-PFC simulation results shown in Fig. 2 of the main text, where $R_c$ is the critical radius for the occurrence of grain instability. Here we show that the standard method of energy minimization is no longer valid for estimating the scaling behavior of this threshold radius $R_c$, which further demonstrates the large degree of deviation between the out-of-equilibrium, nonpotential system studied here and the near-equilibrium, potential counterparts. Consider a 2D crystalline area $\mathcal{A}$ of totally $\mathcal{N}$ grains, subjected to the constraint $\mathcal{N}\pi R^2 = \mathcal{A}$ with variable $R$ and $\mathcal{N}$. If using the standard energetics argument, the rotating crystalline grains are assumed to be governed by a competition between surface line tension $F_{\rm surf}$ and the torsional elastic energy penalty $F_{\rm tor}$. It has been known that \cite{Landau70} the torsional elastic energy density $f_{\rm tor} = \frac{1}{2}C\vartheta_t^2$, where the torsional rigidity $C=\frac{1}{2}\pi\mu R^4$ (with shear modulus $\mu$) for torsional deformation of a cylindrical rod. In the rod a circular cross section of radius $R$ is twisted with respect to the adjoint one by a relative twist angle $d\phi$ and a torsion angle $\vartheta_t = d\phi/dz$ per unit rod length $z$. This can be matched to our case here, where a time series of rotating 2D circular grains can be considered as a time cylinder with each time cross section representing a circular grain state at that time and being rotated with respect to the grain state of the adjoint time step. Thus the rod length $z \rightarrow t$ and torsion angle $\vartheta_t = d\phi/dz \rightarrow \omega = d\phi/dt$, i.e., the rotation frequency with the relative rotation angle $d\phi$ between consecutive time sections. This leads to $f_{\rm tor} = \frac{1}{4}\pi\mu \omega^2 R^4$ for a self-rotating circular grain with radius $R$. Note that the grain rotation rate $\omega$ increases with the net transverse force on the free surface and hence with the increase of $\alpha_1$, but decreases with larger grain size, i.e., $\omega \propto \alpha_1/N^s \sim \alpha_1/R^{2s}$ with $s>0$. Results from both numerical simulations of T-PFC and experimental measurement of starfish embryos rotating clusters (with data obtained from Ref.~\cite{TanNature22}) show that $s \simeq 1$, as seen in Fig.~1B of the main text and Eq.\ \ref{eq:omega}. This yields $f_{\rm tor} = C_0 \alpha_1^2 R^{4-4s}$ where $C_0$ is proportional to the elastic modulus of the crystallite.

The total free energy of the crystalline grains is thus given by 
\begin{equation}
    F_{\rm g} = F_{\rm bulk} + F_{\rm surf} + F_{\rm tor} = \mathcal{N} (f_0 \pi R^2 + \gamma 2\pi R + C_0 \alpha_1^2 R^{4-4s}),
\end{equation}
where $f_0$ is the bulk free energy density and $\gamma$ is the 2D surface or line tension. The free energy variation of those self-rotating grains, $\Delta F_{\rm g} = F_{\rm g} - F_{\rm bulk} = F_{\rm g} - f_0 \mathcal{A}$, is then written as
\begin{equation}
    \Delta F_{\rm g} = \frac{2\mathcal{A}\gamma}{R} + \frac{\mathcal{A} C_0}{\pi} \alpha_1^2 R^{2-4s},
\end{equation}
the minimization of which gives the optimal grain radius in terms of energetics,
\begin{equation}
    R_{\rm opt} = \left [ \frac{\pi\gamma}{C_0(1-2s)} \right ]^{\frac{1}{3-4s}} |\alpha_1|^{-\frac{2}{3-4s}} \sim |\alpha_1|^{-\beta}.
    \label{eq:R_opt}
\end{equation}
Noting that due to the contribution of transverse interaction on the surface tangential force, $\gamma$ would increase with $\alpha_1$ (as seen in Ref.~\cite{CaporussoPRL24} for particle-based simulations of chiral fluid surface), we assume $\gamma \propto \alpha_1^{\delta_\gamma}$ with $\delta_\gamma > 0$.
The scaling exponent for $R_{\rm opt}$ in Eq.\ \ref{eq:R_opt} then becomes 
\begin{equation}
    \beta = \frac{2-\delta_\gamma}{3-4s}.
\end{equation}
Since there is no grain instability in the absence of transverse interaction, $R_{\rm opt} \rightarrow \infty$ as $\alpha_1 \rightarrow 0$, such that $\beta > 0$. The positivity of grain radius imposes a constraint $s < 1/2$ in Eq.\ \ref{eq:R_opt}, giving a range of $0 < \beta < 2$. However, data fitting from both T-PFC simulations and starfish embryo experiments yields $s \simeq 1$, as described above and in the main text. In addition, a larger value of scaling exponent $\beta = 2.5 \pm 0.2$ is obtained from T-PFC simulations, with results shown in Fig.~2A of the main text. Both these indicate that the properties of this odd elastic system is not governed by energy minimization, as it is nonpotential and nonrelaxational.

\bibliography{T-PFC_refs}